\documentclass[graybox,vecphys,natbib]{svmult}

\usepackage{mathptmx}       % selects Times Roman as basic font
\usepackage{helvet}         % selects Helvetica as sans-serif font
\usepackage{courier}        % selects Courier as typewriter font
\usepackage{type1cm}        % activate if the above 3 fonts are
\usepackage{makeidx}         % allows index generation
\usepackage{graphicx}        % standard LaTeX graphics tool
\usepackage{multicol}        % used for the two-column index
\usepackage[bottom]{footmisc}% places footnotes at page bottom

\makeindex             % used for the subject index
\def \lesssim {\mathrel{\vcenter
     {\offinterlineskip \hbox{$<$}\hbox{$\sim$}}}}
\def \gtrsim {\mathrel{\vcenter
     {\offinterlineskip \hbox{$>$}\hbox{$\sim$}}}}
\newcommand{\qvec}[1]{\textbf{\textit{#1}}}

\begin{document}
\bibliographystyle{spbasic}

\title*{Neutrino Emission from Supernovae}
\label{Chapter:NeutrinoEmissionfromSupernovae}
\author{Hans-Thomas Janka}
\institute{Hans-Thomas Janka \at Max Planck Institute for Astrophysics, 
Karl-Schwarzschild-Str.~1, 85748 Garching, Germany\\ \email{thj@mpa-garching.mpg.de}
%\and Name of Second Author \at Name, Address of Institute \email{name@email.address}
}
\maketitle

\abstract*{Supernovae are the most powerful cosmic sources of
MeV neutrinos. These elementary particles
play a crucial role when the evolution
of a massive star is terminated by the collapse of its core
to a neutron star or a black hole and the star explodes as 
supernova. The release of electron neutrinos, which are abundantly 
produced by electron captures, accelerates the catastrophic infall 
and causes a gradual neutronization of the stellar plasma by
converting protons to neutrons as dominant constituents of neutron
star matter.
The emission of neutrinos and antineutrinos of all flavors carries 
away the gravitational binding energy of the compact remnant and
drives its evolution from the hot initial to the cold final state.
The absorption of electron neutrinos and antineutrinos
in the surroundings of the newly formed neutron star can power
the supernova explosion and determines the conditions in the 
innermost supernova ejecta, making them an interesting site for
the nucleosynthesis of iron-group elements and trans-iron nuclei.
In this Chapter the basic neutrino physics in supernova cores and
nascent neutron stars will be discussed. This includes the most
relevant neutrino production, absorption, and scattering processes,
elementary aspects of neutrino transport in dense environments,
the characteristic neutrino emission phases with their typical 
signal features, and the perspectives connected to a measurement 
of the neutrino signal from a future galactic supernova.
}

\abstract{Supernovae are the most powerful cosmic sources of
MeV neutrinos. These elementary particles 
play a crucial role when the evolution
of a massive star is terminated by the collapse of its core
to a neutron star or a black hole and the star explodes as
supernova. The release of electron neutrinos, which are abundantly
produced by electron captures, accelerates the catastrophic infall
and causes a gradual neutronization of the stellar plasma by
converting protons to neutrons as dominant constituents of neutron
star matter.
The emission of neutrinos and antineutrinos of all flavors carries
away the gravitational binding energy of the compact remnant and
drives its evolution from the hot initial to the cold final state.
The absorption of electron neutrinos and antineutrinos
in the surroundings of the newly formed neutron star can power
the supernova explosion and determines the conditions in the
innermost supernova ejecta, making them an interesting site for
the nucleosynthesis of iron-group elements and trans-iron nuclei.
\newline\indent
In this Chapter the basic neutrino physics in supernova cores and
nascent neutron stars will be discussed. This includes the most
relevant neutrino production, absorption, and scattering processes,
elementary aspects of neutrino transport in dense environments,
the characteristic neutrino emission phases with their typical
signal features, and the perspectives connected to a measurement
of the neutrino signal from a future galactic supernova.
}

\section{Introduction}
\label{sec:janka-intro}

The paramount importance of neutrinos in the context of stellar 
core collapse and the question how massive stars achieve to produce
supernova (SN) explosions was first pointed out in seminal papers
by \citet{Colgate1966} and \citet{Arnett1966}. They recognized that the
huge gravitational binding energy\index{gravitational binding energy} 
of a neutron star is carried away by neutrinos, which are therefore 
a copious reservoir of energy for the explosion. Approximating the
neutron star of mass $M_\mathrm{ns}$ and radius $R_\mathrm{ns}$
by a homogeneous sphere with Newtonian gravity, its binding energy,
which roughly equals its gravitational energy, can be estimated as
\begin{equation}
E_\mathrm{b} \sim E_\mathrm{g} \approx 
\frac{3}{5}\,\frac{G M_\mathrm{ns}^2}{R_\mathrm{ns}} 
\approx 3.6\times 10^{53}\,\left(\frac{M_\mathrm{ns}}{1.5\,M_\odot}\right)^{\! 2}
\left(\frac{R_\mathrm{ns}}{10\,\mathrm{km}}\right)^{\! -1}\,\,\mathrm{erg}\,.
\label{eq:ebind}
\end{equation}
If only a fraction of this
energy can be transferred to the gas surrounding the newly formed
neutron star, the overlying stellar layers could be accelerated 
and expelled in a violent blast wave. A major revision of the 
theoretical picture of neutrino effects in collapsing stars 
became necessary after weak neutral 
currents\index{weak neutral currents}, which had 
been predicted in theoretical work by Weinberg and Salam, were 
experimentally confirmed in the early 1970's \citep{Freedman1977}.
With neutral-current scatterings\index{neutral-current scattering} 
of neutrinos off nuclei and free nucleons being possible, it was
recognized that the electron neutrinos, $\nu_e$, produced by 
electron captures can escape freely only at the beginning
of stellar core collapse (which starts out at a density around 
$10^{10}$\,g\,cm$^{-3}$), but get
trapped\index{neutrino trapping} to be carried 
inward with the infalling stellar plasma when the density exceeds
a few times $10^{11}$\,g\,cm$^{-3}$. At this time the implosion
has accelerated so much that the remaining collapse time scale
becomes shorter than the outward diffusion time scale of the 
neutrinos, which increases when scatterings become more and more 
frequent with growing density. Shortly afterwards, typically around
$10^{12}$\,g\,cm$^{-3}$, the electron neutrinos equilibrate with
the stellar plasma and fill up their phase space to form a
degenerate Fermi gas\index{Fermi gas}. 
During the remaining collapse until nuclear
saturation density (about $2.7\times 10^{14}$\,g\,cm$^{-3}$)
is reached, and the incompressibility of the nucleonic
matter due to the repulsive part of the nuclear force enables the
formation of a neutron star, the entropy and the lepton number
(electrons plus electron neutrinos) of the infalling gas 
(stellar plasma plus trapped neutrinos) remain essentially 
constant. Since the change of the entropy by electron
captures and $\nu_e$ escape until trapping is modest, it 
became clear that the collapse of a stellar core proceeds nearly
adiabatically \citep[for a review, see][]{Bethe1990}. 

%-----------------------------------------------------------------------
\begin{figure}[b]
\sidecaption
\includegraphics[scale=0.90]{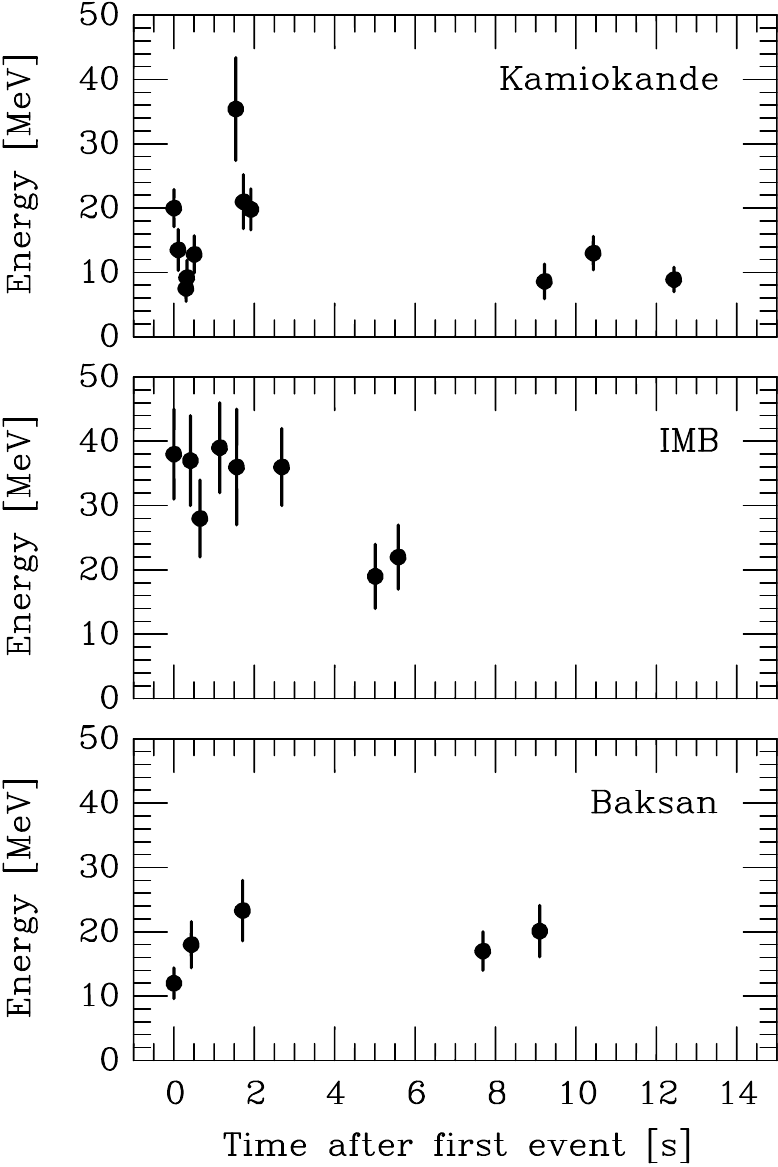}
\caption{Neutrino events recorded by the Kamiokande, IMB and
Baksan underground experiments. The energies do not refer to the
primary electron antineutrinos but to the secondary positrons
produced by the captures of such neutrinos on protons,
$\bar\nu_e + p \longrightarrow n + e^+$. The detector clocks
had unknown relative offsets; while the absolute timing at
IMB had an accuracy of $\pm$50\,ms, the clock at Kamiokande
was accurate only to within $\pm$1\,min, and the time measurement
at Baksan had an uncertainty of $+$2/$-$54\,s. In the
plots the first measured events are synchronized to $t = 0$.
(Figure courtesy of Georg Raffelt)
}
\label{fig:janka-sn1987A}
\end{figure}
%------------------------------------------------------------------------

The proto-neutron star\index{proto-neutron star}, i.e., the hot, 
mass-accreting, still proton- and lepton-rich predecessor
object of the final neutron star, with its super-nuclear densities
and extreme temperatures of up to several $10^{11}$\,K (corresponding
to several 10\,MeV) is highly opaque to all kinds of (active) neutrinos
and antineutrinos. Neutrinos, once generated in this extreme environment,
are frequently re-absorbed, re-emitted, and scattered before they
can reach semi-transparent layers near the ``surface'' of the           
proto-neutron star, which is marked by an essentially exponential 
decline of the density over several orders of magnitude. Before
they finally decouple from the stellar medium closely above this
region and escape, neutrinos have experienced billions of
interactions on average.
The period of time over which the nascent neutron star is able to 
release neutrinos with high luminosities until its gravitational 
binding energy\index{gravitational binding energy}
(Eq.~\ref{eq:ebind}) is radiated away therefore
lasts many seconds \citep{Burrows1986,Burrows1990b}.

%--------------------------------------------------------------------
\begin{figure}[t]
\sidecaption[t]
\includegraphics[scale=0.475,angle=270]{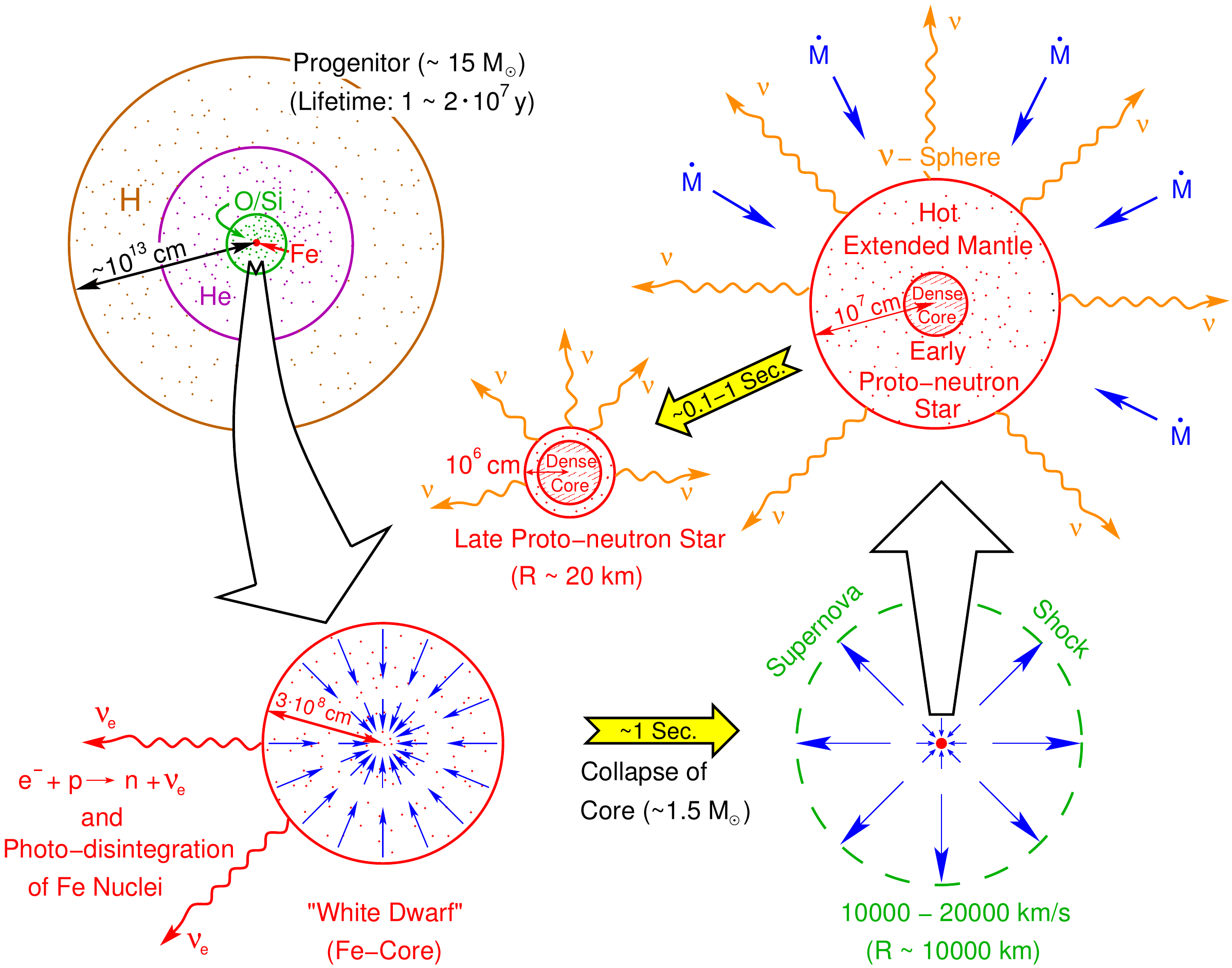}
\caption{Evolution of a massive star from the onset of iron-core
collapse to a neutron star. The progenitor has developed a typical
onion-shell
structure with layers of increasingly heavier elements surrounding
the iron core at the center (upper left corner). Like a white dwarf
star, this iron core (enlarged on the lower left side) is stabilized
mostly by the fermion pressure of nearly degenerate electrons. It
becomes gravitationally unstable when the rising temperatures begin
to allow for partial photo-disintegration of
iron-group nuclei to $\alpha$-particles and nucleons. The contraction
accelerates to a dynamical collapse by electron captures on bound and
free protons, releasing electron neutrinos ($\nu_e$), which initially
escape freely. Only fractions of a second later, the catastrophic infall
is stopped because nuclear-matter density is reached and a proto-neutron
star\index{proto-neutron star} begins to form. 
This gives rise to a strong shock wave which
travels outward and disrupts the star in a supernova explosion
(lower right). The nascent neutron star is initially very extended
(enlarged in the upper right corner), and contracts to a more compact
object while accreting more matter (visualized by the mass-accretion
rate $\dot M$) within the first second of its evolution. This phase as
well as the subsequent cooling and neutronization of the compact remnant
are driven by the emission of neutrinos and antineutrinos of all flavors
(indicated by the symbol $\nu$), which diffuse out from the dense and hot
super-nuclear core over tens of seconds.
\citep[Figure adapted from][]{Burrows1990}
}
\label{fig:janka-nuemission}
\end{figure}
%--------------------------------------------------------------------
 
This expectation was splendidly confirmed by the first and so far 
only detection of neutrinos from a stellar collapse on February 23,
1987, in the case of SN~1987A in the Large Magellanic Cloud at a
distance of roughly 50\,kpc \citep{Raffelt1996}. 
The two dozen neutrino events in the 
three underground experiments of Kamiokande~II \citep{Hirata1987},
Irvine-Michigan-Brookhaven \citep[IMB;][]{Bionta1987}, and
Baksan \citep{Alexeyev1988} were recorded over a time interval
of about 12 seconds (Fig.~\ref{fig:janka-sn1987A}). Also their 
individual energies (up to
40\,MeV) and the associated integrated energy of the neutrino 
signal (some $10^{53}$\,erg) were in the ballpark of model 
predictions and evidenced the birth of a neutron star in this
supernova. Figure~\ref{fig:janka-nuemission} displays a schematic 
representation of the neutrino emission that drives
the evolution from the onset of stellar core collapse to the
cooling of the nascent neutron star, finally leading to a 
neutrino-transparent neutron star with central temperature
below about 1\,MeV (roughly $10^{10}$\,K) within some tens
of seconds.

The neutrino-interaction processes and basic physics
of neutrino transport in supernova matter will be described 
in Sect.~\ref{sec:janka-basics}, the neutrino-emission phases and
corresponding neutrino effects in Sect.~\ref{sec:janka-phases}, 
and the neutrino-emission properties during the different phases
in Sect.~\ref{sec:janka-properties}. Conclusions and an outlook 
will follow in Sect.~\ref{sec:janka-conclusions}.

%-------------------------------------------------------------------------------
\begin{table}
\caption{Most important neutrino processes in supernova and proto-neutron star matter.}
\label{tab:janka-nuprocesses}    
\begin{tabular}{p{7.5cm}p{4.0cm}}
\hline\noalign{\smallskip}
Process &  Reaction$^a$ \\
\noalign{\smallskip}\svhline\noalign{\smallskip}
{\bf Beta-processes (direct URCA processes)} &  \\
electron and $\nu_e$ absorption by nuclei & $e^- + (A,Z) \longleftrightarrow (A,Z-1) + \nu_e$ \\
electron and $\nu_e$ captures by nucleons & $e^- + p \longleftrightarrow  n + \nu_e$  \\
positron and $\bar\nu_e$ captures by nucleons & $e^+ + n \longleftrightarrow  p + \bar\nu_e$ \\
{\bf ``Thermal'' pair production and annihilation processes} &   \\
Nucleon-nucleon bremsstrahlung & $N + N \longleftrightarrow  N + N + \nu + \bar\nu$ \\
Electron-position pair process & $e^- + e^+ \longleftrightarrow \nu + \bar\nu$ \\ 
Plasmon pair-neutrino process  & $\widetilde{\gamma} \longleftrightarrow \nu + \bar\nu$ \\
{\bf Reactions between neutrinos} & \\
Neutrino-pair annihilation & $\nu_e + \bar\nu_e \longleftrightarrow \nu_x + \bar\nu_x$ \\
Neutrino scattering        & $\nu_x + \{\nu_e,\bar\nu_e\} \longleftrightarrow \nu_x + \{\nu_e,\bar\nu_e\}$ \\{\bf Scattering processes with medium particles} & \\
Neutrino scattering with nuclei & $\nu + (A,Z) \longleftrightarrow \nu + (A,Z)$ \\
Neutrino scattering with nucleons & $\nu + N \longleftrightarrow \nu + N$ \\
Neutrino scattering with electrons and positrons & $\nu + e^\pm \longleftrightarrow \nu + e^\pm$ \\
\noalign{\smallskip}\hline\noalign{\smallskip}
\end{tabular}
$^a$ $N$ means nucleons, i.e., either $n$ or $p$, 
$\nu \in \{\nu_e,\bar\nu_e,\nu_\mu,\bar\nu_\mu,\nu_\tau,\bar\nu_\tau\}$,
$\nu_x \in \{\nu_\mu,\bar\nu_\mu,\nu_\tau,\bar\nu_\tau\}$\\
\end{table}
%----------------------------------------------------------------------------------

%------------------------------------------------------------------------
\begin{figure}[t]
\sidecaption[t]
\includegraphics[scale=.45]{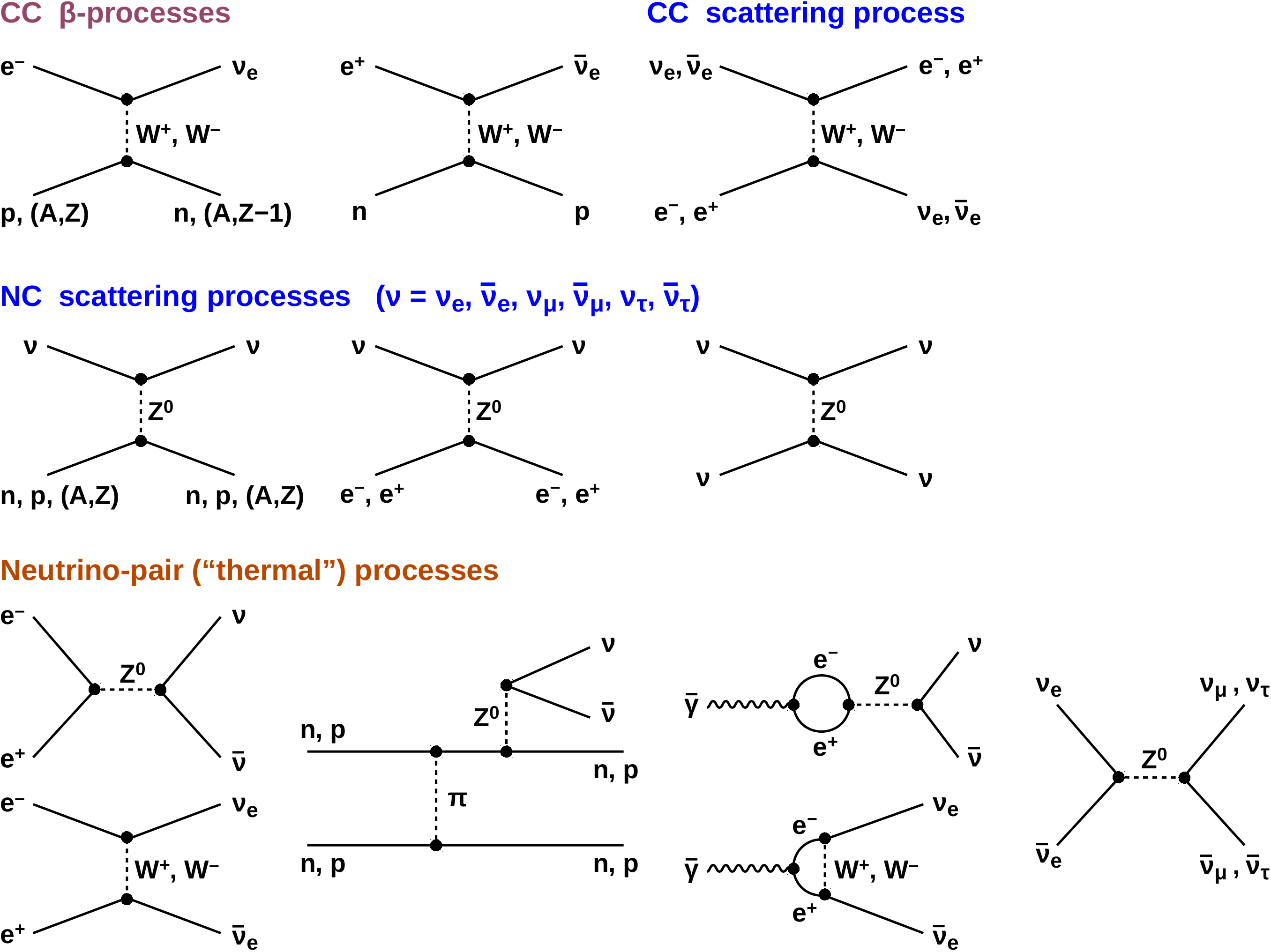}
\caption{Feynman diagrams for the lowest-order contributions to the
most relevant neutrino interactions in supernova cores.
Charged-current (CC) reactions are mediated by $W^\pm$ bosons,
neutral-current (NC) reactions by electrically neutral $Z^0$
bosons. The charged-current 
$\beta$-processes \index{charged-current processes} are responsible
for the production and absorption of $\nu_e$ and $\bar\nu_e$ by
lepton-capture reactions on nucleons ({\em top row, left}).
Scattering processes include the charged-current interactions
of $\nu_e$ and $\bar\nu_e$ with electrons and positrons ({\em top
row, right}) and neutral-current
scatterings\index{neutral-current scatterings} of
neutrinos and antineutrinos of all flavors with nuclei, neutrons,
protons, electrons, positrons, and neutrinos ({\em middle row}).
Neutrino-pair processes are responsible for the creation and
annihilation of neutrino-antineutrino pairs of all flavors.
They include electron-position pair annihilation
through neutral and charged currents, nucleon bremsstrahlung,
the charged- and neutral-current plasmon-neutrino processes,
and neutrino-pair conversion between different flavors ({\em bottom
row, from left to right}).
}
\label{fig:janka-feynman}
\end{figure}
%-----------------------------------------------------------------------

\section{Neutrino Production and Propagation in Supernova Cores}
\label{sec:janka-basics}

In collapsing stars neutrinos and antineutrinos of all flavors
are produced and absorbed by a variety of processes, and,
once created, they scatter off the target particles contained by
the stellar medium as well as off neutrinos, whose number 
densities can exceed those of nucleons and charged leptons 
in some regions of the supernova core. The most important 
interactions at supernova and proto-neutron star conditions
are summarized in
Table~\ref{tab:janka-nuprocesses} and Fig.~\ref{fig:janka-feynman}.

\subsection{Basic Weak Interaction Theory}

According to the Weinberg-Salam-Glashow theory 
(WSG)\index{Weinberg-Salam-Glashow theory}, the weak force between
fermions is mediated by the exchange of massive vector 
bosons\index{vector boson}, namely
two charged intermediate bosons, $W^+$ and $W^-$, and one neutral 
intermediate boson, the $Z^0$. Since the interaction energies at 
typical supernova and proto-neutron star conditions are much smaller
than the rest-mass energies of the $W$ and $Z$ bosons, the WSG 
Hamiltonian density can be rewritten to an effective four-fermion 
point-interaction V--A Hamiltonian (V stands for the vector part
of the interaction, A for the axial-vector part) of the form
\begin{equation}
{\cal H}_\mathrm{weak} = \frac{G_\mathrm{F}}{\sqrt{2}}\,J_\mu^\dagger J^\mu \,,
\label{eq:hweak}
\end{equation}
where $J_\mu$ is the 4-current density of the interacting fermions
and $G_\mathrm{F}$ is the universal Fermi coupling constant,
$G_\mathrm{F} = 1.16637\times 10^{-5}\,\mathrm{GeV}^{-2} = 
1.43588\times 10^{-49}\,$erg\,cm$^3$ (for unit convention of
Planck's constant $\hbar = h/(2\pi) = 1$ and speed of light $c = 1$). 
To lowest non-vanishing order the matrix element of the
interaction\index{matrix element of weak interaction},
${\cal M}$, becomes \citep[e.g.][]{Tubbs1975,Bruenn1985}:
\begin{equation}
{\cal M}(f + \nu \to f' + \nu') = \frac{G_\mathrm{F}}{\sqrt{2}}\,
\overline{\psi}_{f'}\gamma_\mu(C_V-C_A\gamma_5)\psi_f\,
\overline{\psi}_{\nu'}\gamma^\mu(1-\gamma_5)\psi_\nu \,.
\label{eq:mweak}
\end{equation}
Thus expressing low-energy scattering reactions by an effective
neutral-current interaction includes a Fierz-transformed
contribution from $W$ exchange when $f$ is a charged lepton
and $\nu$ the corresponding neutrino. In the case of
charged-current electron and positron captures and
the inverse $\nu_e$ and $\bar\nu_e$ absorptions, $f$ and
$f'$ denote the incoming and outgoing nucleons and  
$\nu$ and $\nu'$ the neutrino and charged lepton in the 
initial and final states. The compound effective coupling coefficients
for the interaction matrix element of Eq.~(\ref{eq:mweak}) are listed
in Table~\ref{tab:janka-coupling} \citep[see also][]{Raffelt2012}.

%--------------------------------------------------------------------------------
\begin{table}
\caption{Effective coupling coefficients in the weak interaction
matrix element of Eq.~(\ref{eq:mweak}). For the effective weak mixing
angle a value of $\sin^2\theta_\mathrm{W}= 0.23146$ was used$^a$.}
\label{tab:janka-coupling}  
\begin{tabular}{p{2.60cm}p{2.5cm}p{2.5cm}p{1.5cm}p{1.25cm}p{0.75cm}}
\hline\noalign{\smallskip}
Fermion $f\,(f')$ & Neutrino (Lepton) & $C_V$ & $C_A$ & $C_V^2$ & $C_A^2$ \\
\noalign{\smallskip}\svhline\noalign{\smallskip}
electron           & $\nu_e$            & $+1/2+2\sin^2\theta_\mathrm{W}$ & $+1/2$    & 0.9272 & 0.25 \\
                   & $\nu_{\mu,\tau}$   & $-1/2+2\sin^2\theta_\mathrm{W}$ & $-1/2$    & 0.0014 & 0.25 \\
proton             & $\nu_{e,\mu,\tau}$ & $+1/2-2\sin^2\theta_\mathrm{W}$ & $+1.26/2$ & 0.0014 & 0.40 \\
neutron            & $\nu_{e,\mu,\tau}$ & $-1/2$                          & $-1.26/2$ & 0.25   & 0.40 \\
neutrino ($\nu_a$) & $\nu_a$            & $+1$                            & $+1$      & 1.00   & 1.00 \\
                   & $\nu_{b\neq a}$    & $+1/2$                          & $+1/2$    & 0.25   & 0.25 \\
neutron (proton)   & $\nu_e$ (electron) & $+1.00$                         & $+1.26$   & 1.00   & 1.59 \\
proton (neutron)   & $\bar\nu_e$ (positron) & $+1.00$                     & $+1.26$   & 1.00   & 1.59 \\
\noalign{\smallskip}\hline\noalign{\smallskip}
\end{tabular}
$^a$ For neutrinos interacting with the same flavor, a factor 2 for an exchange amplitude
for identical fermions was applied. Possible strange-quark contributions to the nucleon spin
were not taken into account for neutral-current neutrino-nucleon scattering. 
\citep[Table adapted from][]{Raffelt2012}
\end{table}
%----------------------------------------------------------------------------------------

With the matrix element being provided by Eq.~(\ref{eq:mweak}), 
the reaction rate, ${\cal R}$, of a neutrino of
energy $q_0$ results from integrating the quantity $(\sigma u)$ 
(having dimensions of cm$^3$\,s$^{-1}$)
over the initial states of the target particle ($\hbar = c = 1$):
\begin{equation}
{\cal R} = \int\frac{\mathrm{d}^3p}{(2\pi)^3}\, {\cal F}(p_0) (\sigma u) \,,
\label{eq:rate}
\end{equation}
where $(\sigma u)$ is the integral over final momentum states of the 
squared matrix element, summed over final spins and averaged over initial spins:
\begin{equation}
\sigma u \equiv \frac{(2\pi)^{-2}}{2p_02q_0} \int\frac{\mathrm{d}^3p'}{2p_0'}
[1-{\cal F}(p_0')] \int\frac{\mathrm{d}^3q'}{2q_0'} [1-{\cal F}(q_0')] 
\left( \frac{1}{2}\sum_\mathrm{spins}|{\cal M}|^2 \right) \delta^4(p+q-p'-q') 
\label{eq:sigmau}
\end{equation}
\citep{Tubbs1975,Burrows2006}. Here, $q$ and $q'$ are the four-momenta of 
the incoming and outgoing lepton, respectively, $p$ and $p'$ the four-momenta
of the interacting fermions in the initial and final states, $q_0$, $q_0'$,
$p_0$, and $p_0'$ the positive time components (energies) of the four-momenta,
and ${\cal F}(E)$ the phase-space occupation functions of fermions of energy 
$E$. While the medium particles are in equilibrium and their phase-space
occupation is described by Fermi-Dirac
distributions\index{phase-space distribution}, the neutrino distribution
can be arbitrary. The magnitude of weak interactions is determined by the 
reference values for the reaction rate and cross section given by
\begin{eqnarray}
{\cal R}_0 &\equiv& 
\frac{\pi}{2}\,c\,\left(\frac{m_ec^2}{2\pi\hbar c}\right)^{\! 3}\sigma_0 
= 3.297\times 10^{39}\sigma_0 \,\,\,\mathrm{cm}^{-2}\,\mathrm{s}^{-1}\,,
\label{eq:r0} \\
\sigma_0 &\equiv& 
\frac{4}{\pi}\,G_\mathrm{F}^2\,\frac{(m_ec^2)^2}{(\hbar c)^4}
= 1.761\times 10^{-44}\,\,\,\mathrm{cm}^2 \,,
\label{eq:sigma0}
\end{eqnarray}
respectively.
Since the squared matrix element is independent of energy, the phase
space integration yields a quadratic dependence of the weak 
interaction cross sections on the particle energy to leading order,
\begin{equation}
\sigma(E) \propto \sigma_0\,\left(\frac{E}{m_ec^2}\right)^{\! 2} \,.
\label{eq:sigma}
\end{equation}

Because of this strong energy dependence of weak interactions,
high-energy neutrinos react much more frequently with medium particles
by scattering and absorption processes and therefore decouple from the stellar
background at a lower density than neutrinos with lower energies.
The mean free path\index{mean free path} between two interactions,
is given by $\lambda(E) = (\sum_i n_{\mathrm{t},i}\,\sigma_i(E))^{-1} 
\equiv (\rho\kappa_\mathrm{tot}(E))^{-1}$, 
where $n_{\mathrm{t},i}$ is the number density 
of target particles of species $i$, $\sigma_i(E)$ the corresponding 
interaction cross section with neutrinos,
$\rho$ the matter density, and $\kappa_\mathrm{tot}(E)$ the total
opacity\index{opacity} in units of cm$^2$\,g$^{-1}$. When $\lambda(E)$ 
includes contributions from all neutrino
interactions, neutrino decoupling takes place at the neutrinospheric
radius $R_\nu(E)$ defined as the radial position where the optical
depth\index{optical depth} is unity:
\begin{equation}
\tau_\nu(E) =
\int_{R_\nu(E)}^\infty \frac{\mathrm{d}r}{\lambda(r,E)} = 
\int_{R_\nu(E)}^\infty \mathrm{d}r\,\,\rho(r)\kappa_\mathrm{tot}(r,E) 
= 1\,.
\label{eq:neutrinosphere}
\end{equation}
Frequent scatterings as well as absorption and re-emission induce 
a random-walk motion of the neutrinos on their way out of the 
deep interior to the neutrino transparent regime at low densities.
Over a (small) 
vertical distance $\Delta z$ to the surface, neutrinos of energy
$E$ experience an average number of $N_\mathrm{ia}$ collisions with
target particles, which is given by the relation:
\begin{equation}
N_\mathrm{ia}^{1/2}\lambda(E) = \Delta z \sim \tau_\nu(E)\lambda(E) \,.
\label{eq:numberia}
\end{equation}
The neutrinosphere\index{neutrinosphere} at $\tau_\nu(E) = 1$ is 
therefore defined as the location where neutrinos of energy $E$ undergo
on average one final interaction, $N_\mathrm{ia} = 1$, prior to escape.

\subsection{Neutrino Transport}

Neutrino transport in supernova cores involves a diffusive mode 
of propagation at the high densities of the newly formed neutron
star, a gradual and energy-dependent decoupling of the neutrinos 
in the neutrinospheric region,
and the transition to free streaming when the neutrinos escape 
from the neutron star. The evolution of the neutrino phase-space 
distribution function\index{phase-space distribution} 
${\cal F}(\qvec{r},\qvec{q},t)$
in these different regimes is described by the Boltzmann 
transport equation\index{Boltzmann equation} 
\citep[e.g.][]{Burrows2000,Liebendoerfer2001,Liebendoerfer2004,Rampp2002,Mezzacappa2004},
\begin{equation}
\frac{\mathrm{D}{\cal F}(\qvec{r},\qvec{q},t)}{\mathrm{D}t} = 
\frac{\partial{\cal F}}{\partial t} +
\frac{\partial\qvec{r}}{\partial t}\,\nabla_{\qvec{r}}{\cal F} +
\frac{\partial\qvec{q}}{\partial t}\,\nabla_{\qvec{q}}{\cal F} =
{\cal C}({\cal F}) \,,
\label{eq:boltzmann}
\end{equation}
where $\mathrm{D}/\mathrm{D} t$ denotes the total derivative of 
${\cal F}(\qvec{r},\qvec{q},t)$ with 
respect to time $t$. $\nabla_{\qvec{r}}$ and $\nabla_{\qvec{q}}$ are
the partial derivatives with respect to the space coordinates, 
$\qvec{r}$, and momentum coordinates, $\qvec{q} = q\qvec{n}$,
when $\qvec{n}$ defines the unit vector in the direction of neutrino
propagation. On the r.h.s.\ of Eq.~(\ref{eq:boltzmann}), 
${\cal C}({\cal F})$ stands for the
collision integral that contains all rates of neutrino production, 
absorption, annihilation, and scattering processes. Moreover, since
supernova neutrinos possess typical energies in the MeV range, which
is much larger than the experimental rest-mass limit for active
flavors, $m_\nu c^2 < 1$\,eV, they propagate essentially with the
speed of light $c$. Therefore one can use $|\qvec{q}|= q = E/c$ and
$\partial\qvec{r}/\partial t = c\,\qvec{n}$. The momentum derivative
in Eq.~(\ref{eq:boltzmann}), $\partial\qvec{q}/\partial t$, accounts
for the effects of forces on the neutrino, e.g.\ in the form of
gravitational redshifting.

Note that for reasons of simplicity, Eq.~(\ref{eq:boltzmann}) was
written in a flat spacetime. In practice, the solution of this
equation faces a lot of complications not only due to spacetime 
curvature in general relativity. For the most general case, where
non-isoenergetic scattering redistributes neutrinos in 
energy-momentum space, Eq.~(\ref{eq:boltzmann}) is an
integro-differential equation. Final-state fermion blocking and
neutrino-anti\-neutrino coupling in pair processes and 
neutrino-neutrino scattering (Table~\ref{tab:janka-nuprocesses} 
and Fig.~\ref{fig:janka-feynman}) make the problem non-linear in 
${\cal F}$. Moreover, the motion of the stellar fluid has to be
accounted for by Lorentz transformations and requires the choice
of solving for ${\cal F}$ in the comoving frame of the fluid, 
where the collision integral is most easily treated, or in the 
laboratory frame, where the left-hand side of Eq.~(\ref{eq:boltzmann})
retains its simple form. 

With the neutrino phase-space distribution function ${\cal F}$ being
determined as solution of Eq.~(\ref{eq:boltzmann}), the quantities
characterizing the neutrino emission can be computed as integrals
over the coordinates of the momentum vector $\qvec{q} = (E/c)\qvec{n}$.
This yields for the specific neutrino number density, 
$\mathrm{d}n_\nu/\mathrm{d}E$, specific energy density,
$\mathrm{d}\varepsilon_\nu/\mathrm{d}E$, specific number flux, 
$\mathrm{d}\qvec{F}_n/\mathrm{d}E$, specific energy flux, 
$\mathrm{d}\qvec{F}_e/\mathrm{d}E$, and for
the corresponding total number and energy densities and fluxes
(taking into account that there is one spin state of either
left-handed neutrinos or right-handed antineutrinos):
\begin{eqnarray}
\frac{\mathrm{d}n_\nu(\qvec{r},E,t)}{\mathrm{d}E}
&=& \frac{1}{(hc)^3}\,E^2\, \int_{4\pi}\mathrm{d}\Omega\,
{\cal F}(\qvec{r},\qvec{q},t)\,,\label{eq:mom1}\\
\frac{\mathrm{d}\varepsilon_\nu(\qvec{r},E,t)}{\mathrm{d}E}
&=& \frac{1}{(hc)^3}\,E^3\,\int_{4\pi}\mathrm{d}\Omega\,
{\cal F}(\qvec{r},\qvec{q},t)\,,\label{eq:mom2}\\
\frac{\mathrm{d}\qvec{F}_n(\qvec{r},E,t)}{\mathrm{d}E}
&=& \frac{c}{(hc)^3}\,E^2\,\int_{4\pi}\mathrm{d}\Omega\,\qvec{n}\,
{\cal F}(\qvec{r},\qvec{q},t)\,,\label{eq:mom3}\\
\frac{\mathrm{d}\qvec{F}_e(\qvec{r},E,t)}{\mathrm{d}E}
&=& \frac{c}{(hc)^3}\,E^3\,\int_{4\pi}\mathrm{d}\Omega\,\qvec{n}\,
{\cal F}(\qvec{r},\qvec{q},t)\,,\label{eq:mom4}\\
n_\nu(\qvec{r},t) &=& \int_0^\infty\mathrm{d}E\,
\frac{\mathrm{d}n_\nu}{\mathrm{d}E}\,,\quad
\varepsilon_\nu(\qvec{r},t) = \int_0^\infty\mathrm{d}E\,
\frac{\mathrm{d}\varepsilon_\nu}{\mathrm{d}E}\,,\label{eq:mom5}\\
\qvec{F}_n(\qvec{r},t) &=& \int_0^\infty\mathrm{d}E\,
\frac{\mathrm{d}\qvec{F}_n}{\mathrm{d}E}\,,\quad
\qvec{F}_e(\qvec{r},t) = \int_0^\infty\mathrm{d}E\,
\frac{\mathrm{d}\qvec{F}_e}{\mathrm{d}E}\,,\label{eq:mom6}
\end{eqnarray}
where $\mathrm{d}\Omega$ is the solid angle element around unit vector
$\qvec{n}$. The flux through an area with normal unit vector $\qvec{m}$ is
given by $\qvec{F}\qvec{m}$, and the ratio of neutrino flux and neutrino
density yields the so-called flux or streaming factor,
$\qvec{s}_n = \qvec{F}_n/(n_\nu c)$ and 
$\qvec{s}_e = \qvec{F}_e/(\varepsilon_\nu c)$.
The energy moments\index{neutrino spectrum!energy moments}
of order $k$ (i.e., the average values of $E^k$) 
for the local neutrino number
density, $\langle E^k\rangle$, and for the neutrino number flux,
$\langle E^k\rangle_\mathrm{flux}$, are given by:
\begin{eqnarray}
\langle E^k\rangle&=&\int_0^\infty\!\mathrm{d}E\,E^{2+k}\!
\int_{4\pi}\!\mathrm{d}\Omega\,{\cal F}(\qvec{r},\qvec{q},t)\!\cdot\!
\left\{\int_0^\infty\!\mathrm{d}E\,E^2\! \int_{4\pi}\!\mathrm{d}\Omega\,
{\cal F}(\qvec{r},\qvec{q},t)\right\}^{-1}\!\!\!\!\!\!\! ,\\
\langle E^k\rangle_\mathrm{flux} &=&
\left|\int_0^\infty\!\mathrm{d}E\,E^{2+k}\! 
\int_{4\pi}\!\mathrm{d}\Omega\,\qvec{n}{\cal F}(\qvec{r},\qvec{q},t)\right| 
\! \cdot\! \left|\int_0^\infty\!\mathrm{d}E\,E^2\! 
\int_{4\pi}\!\mathrm{d}\Omega\,\qvec{n}{\cal F}(\qvec{r},\qvec{q},t)\right|^{-1}
\!\!\!\!\!\!\! .
\end{eqnarray}
The rms energies of neutrino energy density and neutrino energy flux
are defined as 
$\langle E\rangle_\mathrm{rms} = \sqrt{\langle E^3\rangle/\langle E\rangle}$
and
$\langle E\rangle_\mathrm{rms,flux} = \sqrt{\langle E^3\rangle_\mathrm{flux}/
\langle E\rangle_\mathrm{flux}}$.

Since the solution of the time-dependent Boltzmann equation in
three spatial dimensions with its full energy-momentum dependence
is not feasible on current supercomputers,
a variety of different approximations are applied, for example
by reducing the number of momentum-space variables by one
in the so-called ``ray-by-ray'' approach \citep{Buras2006},
which assumes the neutrino
phase-space distribution\index{phase-space distribution} ${\cal F}$
to be axi-symmetric around one, typically the radial,
direction and thus ignores non-radial flux components.
Alternatively, the dependence of the Boltzmann equation on the
momentum directions can be removed by integration over all
directions after multiplication with different powers of $\qvec{n}$,
by which means an infinite set of so-called moment
equations\index{Boltzmann equation!moment equations} is derived,
in which angular moments\index{phase-space distribution!angular moment} 
(integrals) of ${\cal F}$ (like those of Eqs.~\ref{eq:mom1}--\ref{eq:mom4})
appear as dependent
variables. Because on each level more moments than equations occur,
a termination of the set on any level requires to involve a
closure relation, which in most cases is a chosen function between
the available moments. A termination on the level of the first
moment equation, which is the neutrino energy equation, leads to
the diffusion treatment\index{diffusion approximation}. The 
compatibility of the diffusion flux (which diverges in the transparent
regime) with the causality limit is usually ensured by the use of a 
flux limiter \citep[e.g.][]{Bruenn1985}\index{flux limiter}. 
A termination on the
level of the second moment equation, which is the neutrino momentum
equation, yields the so-called two-moment transport 
approximation\index{two-moment transport approximation}.

%--------------------------------------------------------------------------
\begin{figure}[t]
\sidecaption[t]
\includegraphics[scale=.47]{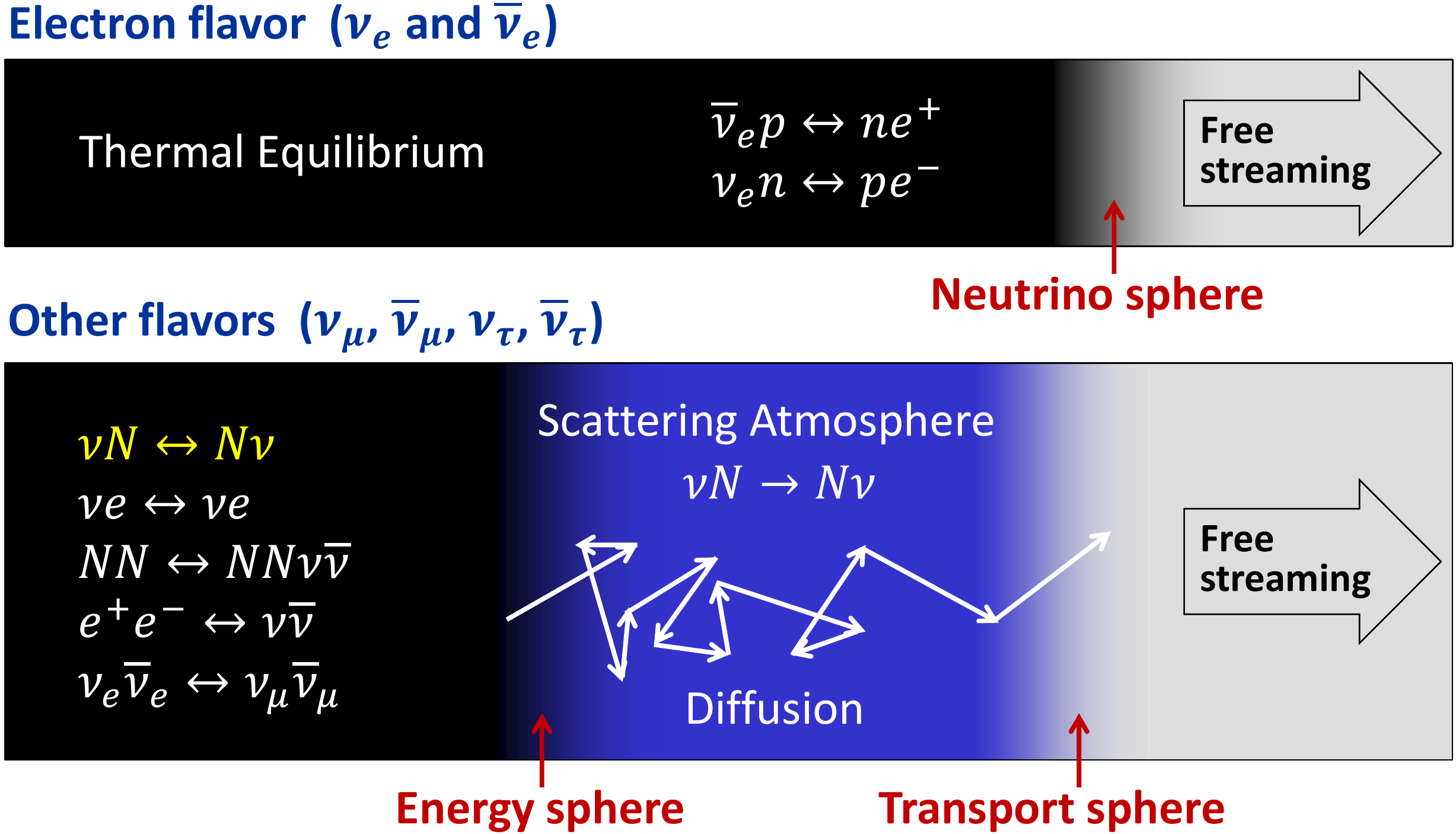}
\caption{Sketch of the transport properties of electron-flavor
neutrinos and antineutrinos ({\em upper part}) compared to
heavy-lepton neutrinos ({\em lower part}). In the supernova core
$\nu_e$ and
$\bar\nu_e$ interact with the stellar medium by charged-current
absorption and emission reactions, which provide a major 
contribution to their opacities
and lead to a strong energetic coupling up to the location of their
neutrinospheres, outside of which both chemical equilibrium between
neutrinos and stellar matter (indicated by the black region) and
diffusion cannot be maintained. In contrast,
heavy-lepton neutrinos are energetically less tightly coupled
to the stellar plasma, mainly by pair creation reactions like
nucleon bremsstrahlung, electron-position annihilation and
$\nu_e\bar\nu_e$ annihilation. The total opacity, however, is
determined mostly by neutrino-nucleon scatterings, whose small energy
exchange per scattering does not allow for an efficient energetic
coupling. Therefore heavy-lepton neutrinos fall out of thermal equilibrium
at an energy sphere that is considerably deeper inside the nascent
neutron star than the transport sphere, where the transition from
diffusion to free streaming sets in. The blue band indicates the
scattering atmosphere where the heavy-lepton neutrinos still collide
frequently with neutron and protons and lose some of their energy,
but cannot reach equilibrium with the background medium any longer.
\citep[Figure adapted from][courtesy of Georg Raffelt]{Raffelt2012}
}
\label{fig:janka-nutrans}
\end{figure}
%---------------------------------------------------------------------------

\subsection{Flavor-dependent Neutrino Decoupling}
\label{sec:janka-decoupling}

Since electrons and positrons are very abundant at the temperatures
in supernova
cores, whereas muons and tauons with their high rest masses are
not, $\nu_e$ and $\bar\nu_e$ interact not only by neutral-current
processes but also via charged-current reactions
(Tables~\ref{tab:janka-nuprocesses} and \ref{tab:janka-coupling}; 
Fig.~\ref{fig:janka-feynman}). This causes distinct differences
of their transport behavior compared to heavy-lepton neutrinos
($\nu_x = \nu_\mu,\,\bar\nu_\mu,\,\nu_\tau,\,\bar\nu_\tau$),
in particular concerning their decoupling near the
neutrinosphere.

Charged-current $\beta$-processes provide a major
contribution to the total opacity of $\nu_e$ and $\bar\nu_e$,
because the interaction cross sections of these reactions are big.
Frequent captures and re-emission of these
neutrinos at the local conditions of
temperature and density are efficient in keeping them
fairly close to local thermodynamic equilibrium (i.e., near
thermal and chemical equilibrium) until they begin their transition
to free streaming at their corresponding energy-averaged neutrinosphere. 
This sphere is also called transport sphere\index{transport sphere}
(sometimes also ``scattering sphere''),
whose radius $R_{\nu,\mathrm{t}}$ is determined by solving
Eq.~(\ref{eq:neutrinosphere}) with a suitable spectral average
of the total opacity\index{total opacity} 
$\kappa_\mathrm{tot}\equiv \kappa_\mathrm{abs}+\kappa_\mathrm{scatt}$,
which includes all contributions from scattering and absorption
processes. Equilibration between neutrinos and the stellar
background is possible up to the so-called average
energy sphere\index{energy sphere} (also termed ``number sphere'',
because outside of this location the number of neutrinos of a
certain species is essentially fixed).
When scatterings increase the zig-zag path of neutrinos
diffusing through the medium and thus increase the probability
of neutrinos to be absorbed, the radius $R_{\nu,\mathrm{e}}$ of
the energy sphere is given by the condition
\begin{equation}
\tau_\mathrm{eff} = \int_{R_{\nu,\mathrm{e}}}^\infty
\mathrm{d}r\,\rho \kappa_\mathrm{eff} = \frac{2}{3} 
\label{eq:esphere}
\end{equation}
\citep{Shapiro1983,Raffelt2001}.
Here, the effective optical depth is defined as
\begin{equation}
\kappa_\mathrm{eff} = \sqrt{\kappa_\mathrm{abs}\kappa_\mathrm{tot}}\ .
\label{eq:effodepth}
\end{equation}
In the integral of Eq.~(\ref{eq:esphere}), again a suitable 
average of $\kappa_\mathrm{eff}$ over the energy spectrum 
has to be used. Since scattering
and absorption contribute roughly equally to the total opacity of 
$\nu_e$ and $\bar\nu_e$, i.e., $\kappa_\mathrm{abs}\approx
(1/2)\kappa_\mathrm{tot}$ and therefore $\kappa_\mathrm{eff} \approx
\kappa_\mathrm{tot}/\sqrt{2}\sim (2/3)\kappa_\mathrm{tot}$, the 
energy and transport sphere turn out to be nearly identical:
$R_{\nu_e,\mathrm{e}} \approx R_{\nu_e,\mathrm{t}}$
and $R_{\bar\nu_e,\mathrm{e}} \approx R_{\bar\nu_e,\mathrm{t}}$
(see Fig.~\ref{fig:janka-nutrans}).

The situation is different for the heavy-lepton species $\nu_x$. 
These are created and destroyed only as neutrino-antineutrino pairs
in neutral-current reactions (cf.\ Fig.~\ref{fig:janka-feynman} and
Table~\ref{tab:janka-nuprocesses}). While at high densities the
main pair-production process is nucleon bremsstrahlung (with
the plasmon-decay process contributing on a secondary level),
electron-positron and $\nu_e\bar\nu_e$ annihilation take over
as the dominant producers of $\nu_x\bar\nu_x$ pairs at densities
below about 10\% of nuclear matter density, where the stellar
medium is less degenerate and larger numbers of positrons and
electron antineutrinos are present. The total opacity of $\nu_x$,
however, is
largely dominated by neutral-current scatterings off nucleons
because of the much greater cross sections of these reactions. 
As a consequence, the average energy sphere of $\nu_x$ is located
at considerably higher density, deeper inside the nascent neutron
star, than their average transport sphere:
$R_{\nu_x,\mathrm{e}} < R_{\nu_x,\mathrm{t}}$ 
(Fig.~\ref{fig:janka-nutrans}).
While outside of the energy sphere the number flux of each 
species of $\nu_x$ is essentially conserved, the energy flux
can still change between the energy and transport spheres
because of energy transfers in the frequent collisions with
nucleons (and to a lesser degree with electrons and 
electron-type neutrinos), in which mostly the energetic
neutrinos from the high-energy tail of the $\nu_x$ spectrum
can deliver energy to the cooler stellar medium
\citep{Raffelt2001,Keil2003}.

The neutral-current nucleon scattering opacities of heavy-lepton 
neutrinos and antineutrinos are to lowest order the same \citep[with 
only minor higher-order differences associated with weak-magnetism
corrections][]{Horowitz2002}.
In the absence of large concentrations of muons, also muon
and tau neutrinos interact with the medium essentially symmetrically.
For these reasons the four species of heavy-lepton neutrinos are
treated as one kind of $\nu_x$ in many applications.

As the neutronization\index{neutronization} and
deleptonization\index{deleptonization} of the nascent
neutron star progress due to the ongoing conversion of 
electrons and protons to neutrons and the escape of $\nu_e$,
the decreasing abundance of protons reduces the
absorption opacity of $\bar\nu_e$. Therefore, as time goes on,
the opacity of electron antineutrinos becomes more and more
similar to that of heavy-lepton (anti)neutrinos, and the 
radiated spectra of these neutrinos match each other closely.

%--------------------------------------------------------------------------
\begin{figure}[!]
\sidecaption[t]
\includegraphics[scale=.25,angle=270]{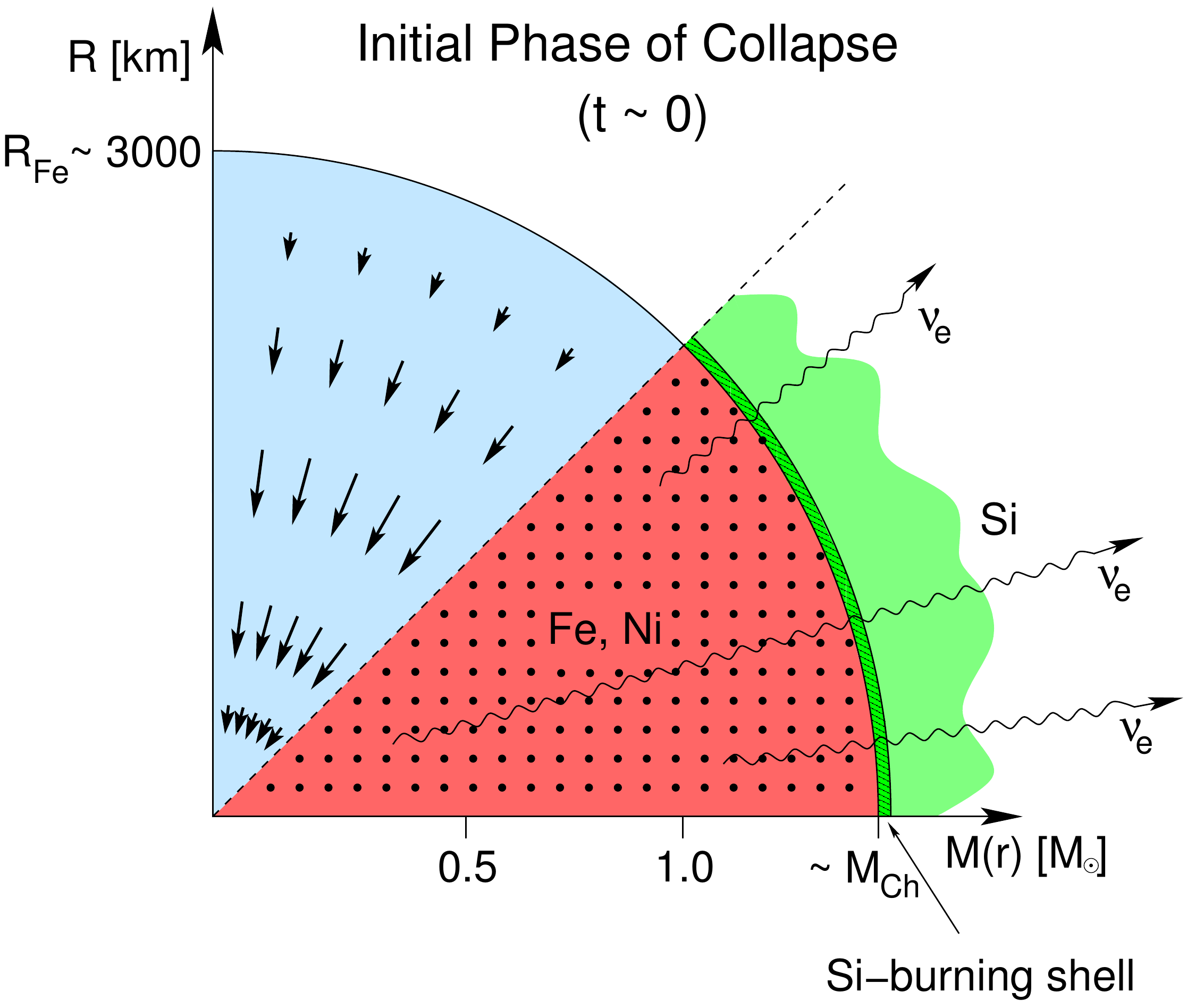}\hspace{25pt}
\includegraphics[scale=.25,angle=270]{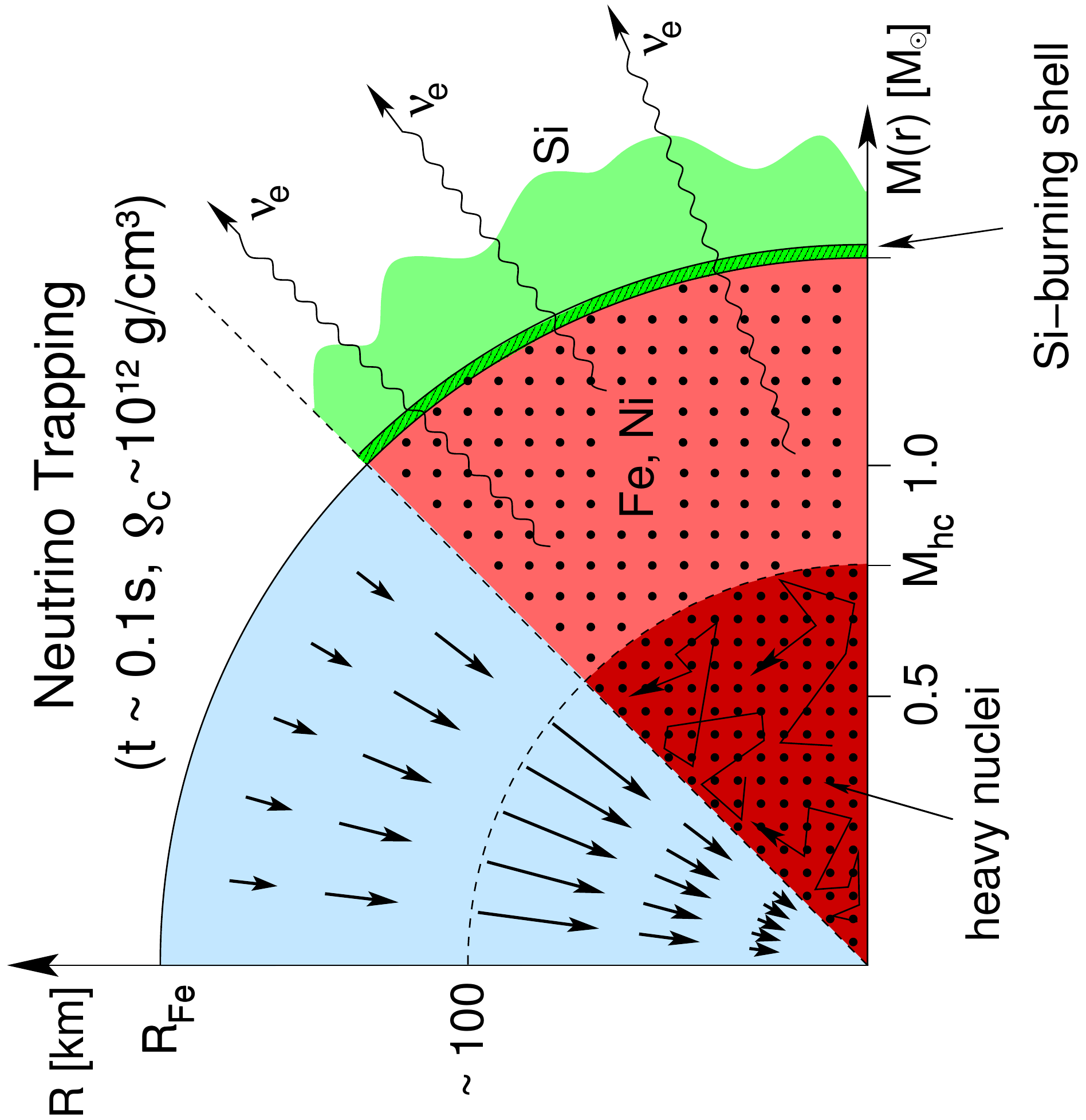}\\
\includegraphics[scale=.25,angle=270]{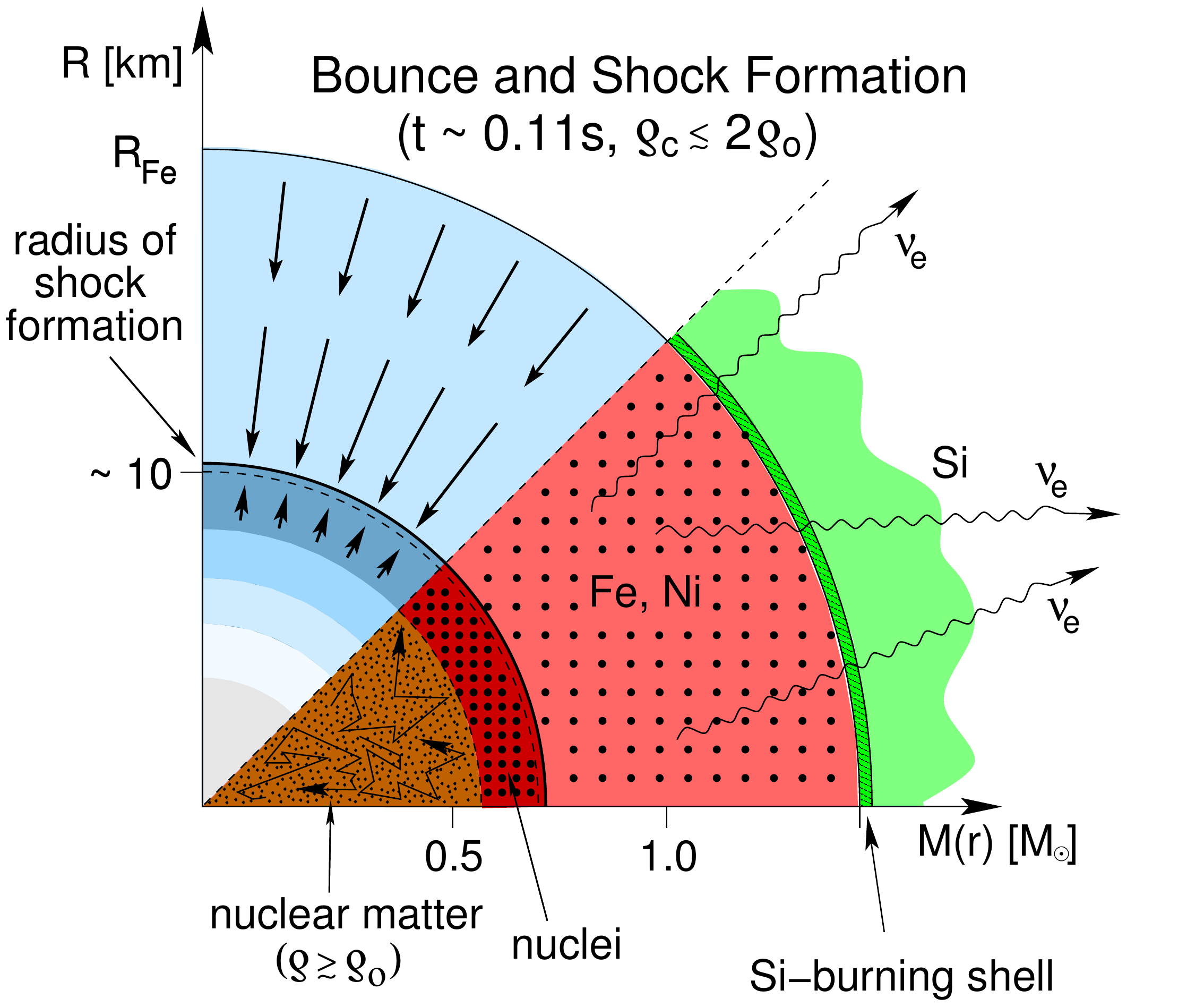}\hspace{10pt}
\includegraphics[scale=.25,angle=270]{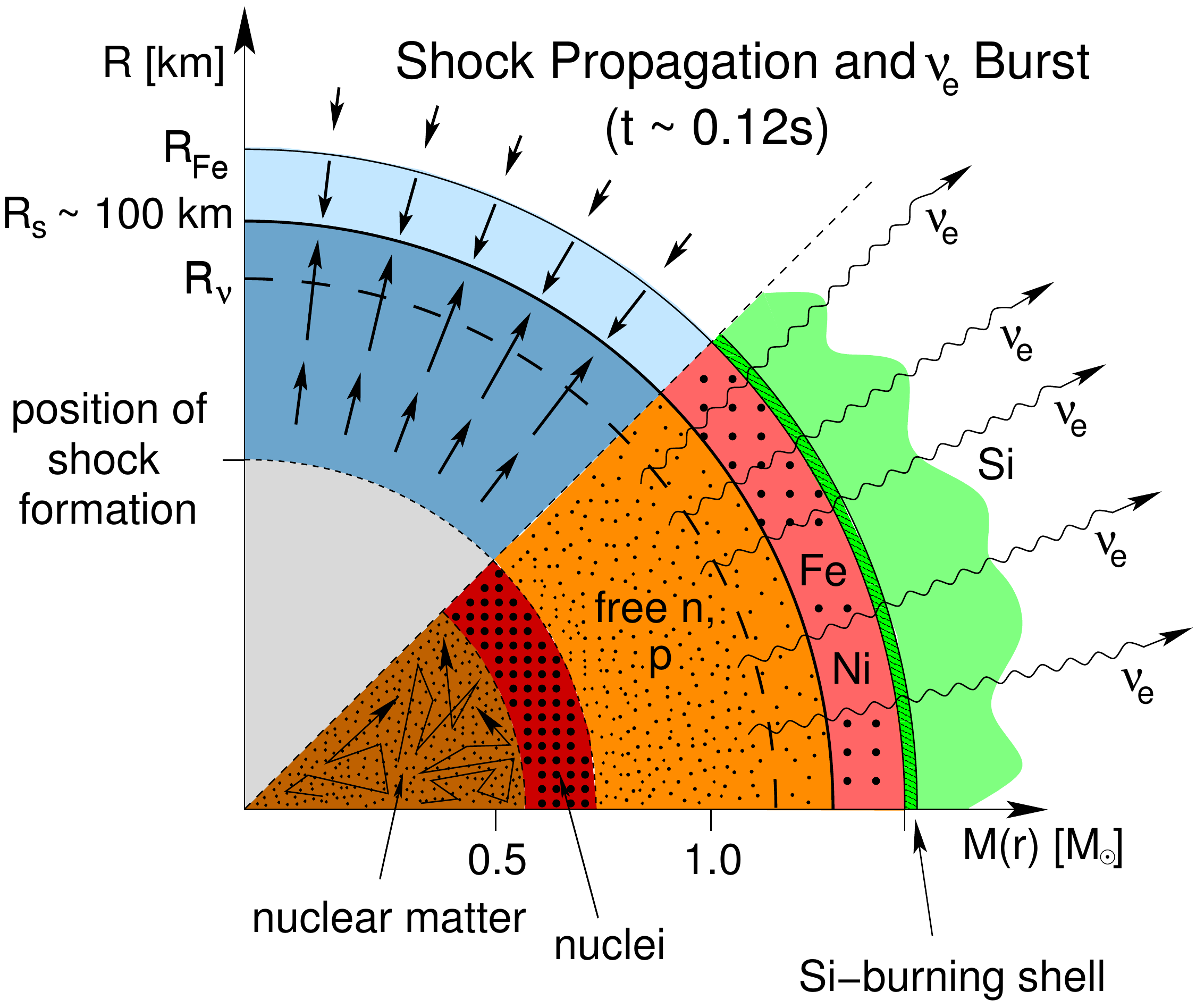}\\
\includegraphics[scale=.25,angle=270]{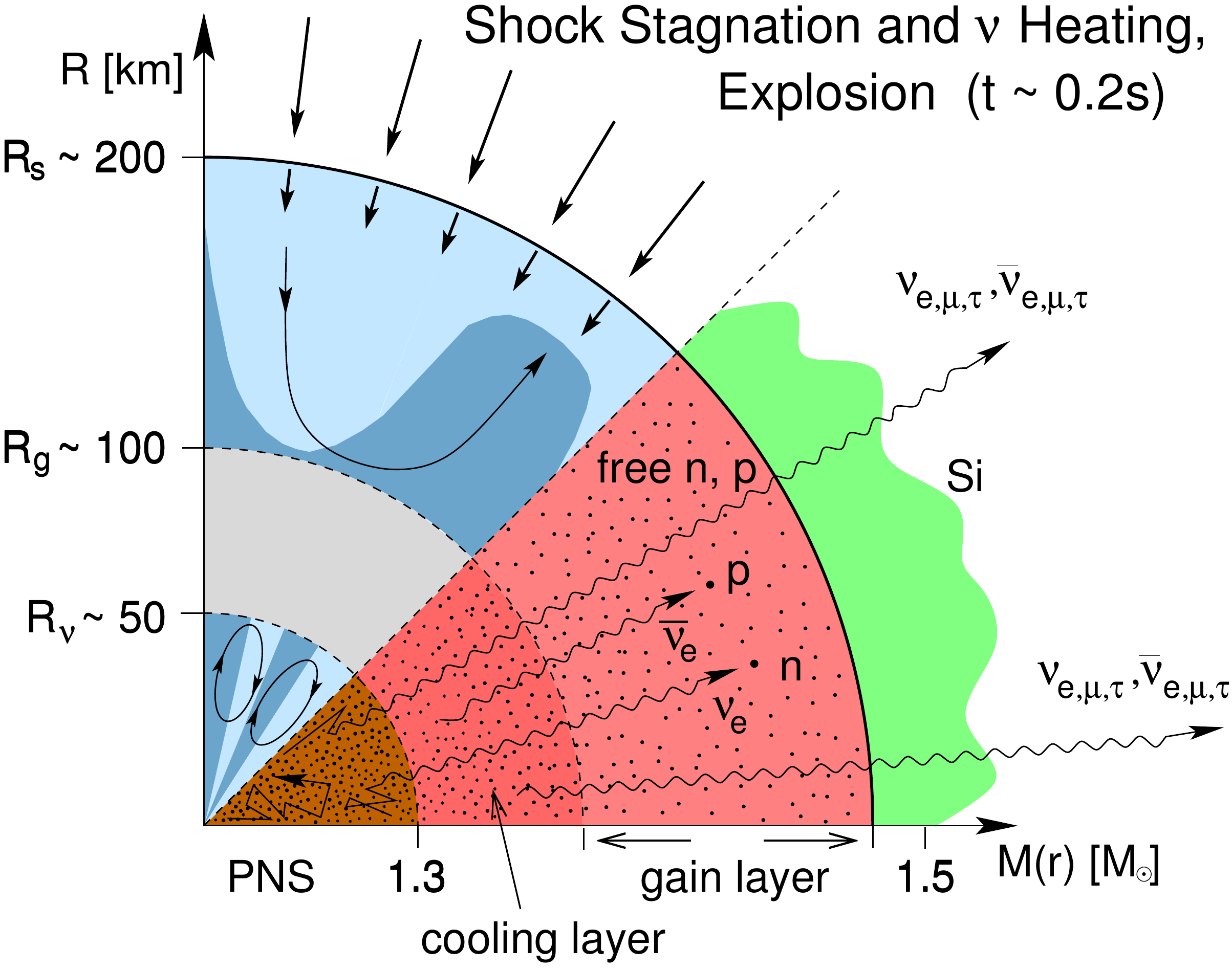}
\includegraphics[scale=.25,angle=270]{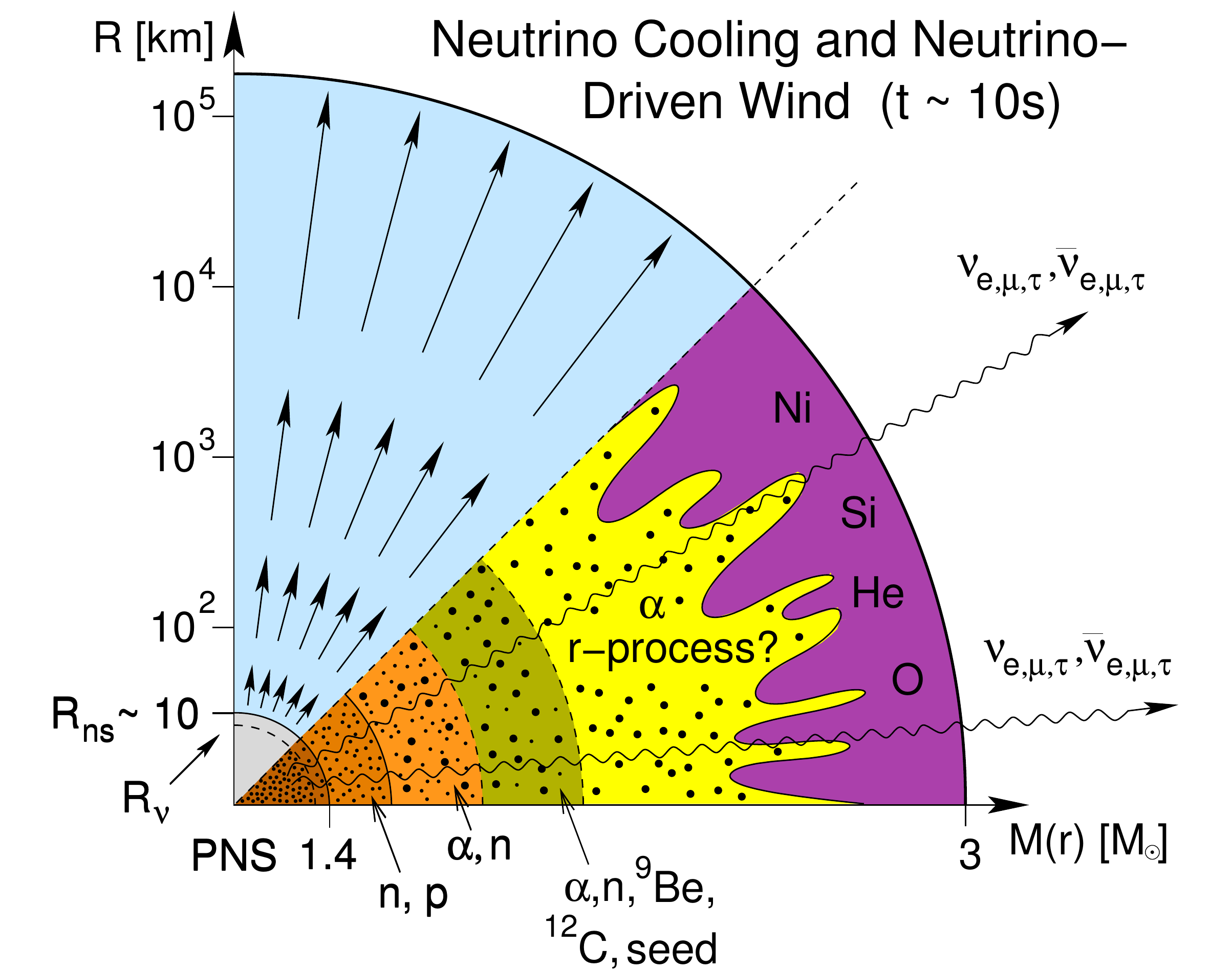}
\caption{Six phases of neutrino production and its
dynamical consequences ({\em from top left to bottom right}). In the
lower halves of the plots the composition of the stellar medium and
the neutrino effects are sketched, while in the
upper halves the flow of the stellar matter is shown by arrows.
Inward pointing arrows denote contraction or collapse, outward
pointing arrows expansion or mass ejection.
Radial distances $R$ are indicated on the vertical axes, the
corresponding enclosed masses $M(r)$ are given on the horizontal axes.
$R_\mathrm{Fe}$, $R_\mathrm{s}$, $R_\nu$, $R_\mathrm{g}$, and
$R_\mathrm{ns}$ denote the iron-core radius, shock radius, neutrinospheric
radius, gain radius (which separates neutrino cooling and 
heating layers), and proto-neutron star (PNS) radius, respectively.
$M_\mathrm{Ch}$ defines the effective Chandrasekhar mass,
$M_\mathrm{hc}$ the mass of the homologously collapsing
inner core\index{homologously collapsing core}
(where velocity $u\propto r$), $\rho_\mathrm{c}$ the central density,
and $\rho_0\approx 2.7\times 10^{14}$\,g\,cm$^{-3}$ the nuclear
saturation density. \citep[Figure taken from][]{Janka2007}
}
\label{fig:janka-nuphases}
\end{figure}
%-----------------------------------------------------------------------------

\section{Neutrino Emission Phases}
\label{sec:janka-phases}

Figure~\ref{fig:janka-nuphases} presents a series of plots 
that provide an overview of the main processes and regions of 
neutrino production and their dynamical effects during the
collapse of a stellar core and on the way to the supernova
explosion. Neutrinos exchange lepton number,
energy and momentum with the stellar medium. Corresponding
source terms must therefore be taken into account in the 
hydrodynamics equations that describe the 
time evolution of the stellar medium in terms of conservation
laws for mass, momentum, energy, and (electron) lepton number.

\runinhead{(1) Onset of Stellar Core Collapse} 
The slow contraction of the growing and aging iron core, which
develops a progenitor-star dependent mass between 
$\sim$1.3\,$M_\odot$ and $\sim$2\,$M_\odot$, speeds up when its
central temperature approaches 1\,MeV ($\approx$\,10$^{10}$\,K).
At this stage, thermal $\gamma$ photons become sufficiently energetic
to partially disintegrate\index{photo-disintegration} the iron-group
nuclei to $\alpha$-particles and free nucleons. This converts
thermal energy to rest-mass energy, thus overcoming the binding
energy of nucleons in the nuclei, and causes a reduction of 
the effective adiabatic index\index{adiabatic index}, i.e., 
of the increase of the pressure with rising density, below the
critical value of $4/3$. (General relativistic effects lead to a 
slight upward correction of this critical value, rotation to a
slight reduction.) Since the Fermi
energy of the degenerate electrons also rises, electron captures on 
nuclei become possible \citep[for the current state-of-the-art 
of the treatment, see][]{Langanke2003,Juodagalvis2010,Balasi2015}. 
Initially, the $\nu_e$ thus produced escape
unimpeded (Fig.~\ref{fig:janka-nuphases}, top left panel).

\runinhead{(2) Neutrino Trapping} 
When the density exceeds a few times $10^{11}$\,g\,cm$^{-3}$,
neutrinos begin to become trapped in the collapsing stellar 
core\index{neutrino trapping}. From this moment on,
the $\nu_e$ produced by ongoing electron captures
---now dominantly on free protons--- are swept inward with
the infalling matter, and entropy as well as lepton number
are essentially conserved in the contracting flow 
(Fig.~\ref{fig:janka-nuphases}, 
top right panel). Neutrino trapping is mainly a
consequence of neutral-current scattering of low-energy
neutrinos on heavy nuclei, whose nucleons act as one coherent
scatterer. Because the vector parts of the neutrino-neutron 
scattering amplitudes dominate compared to those of protons
(cf.\ Table~\ref{tab:janka-coupling})
and add up in phase, whereas the overall axial-vector current
is reduced by spin-pairing of the nucleons in nuclei, the 
coherent scattering cross section
effectively scales with the square of the neutron number $N$:
\begin{equation}
\sigma_{A,\mathrm{coh}}\approx \frac{1}{16}\,\sigma_0\,
\left(\frac{E}{m_e c^2}\right)^{\! 2} N^2 \,.
\label{eq:sigcoh}
\end{equation}
As electrons continue to be converted to $\nu_e$, the dynamical
collapse accelerates to nearly free-fall velocities (up to
$\sim$30\% of the speed of light) in the supersonic outer
core region. The inner core implodes subsonically and 
homologously\index{homologously collapsing core}, i.e.\ with 
a velocity that is proportional to the radius,
which implies a self-similar change of the structure.
The size of the homologous core is roughly given by the 
instantaneous Chandrasekhar mass,
\begin{equation}
M_\mathrm{hc} \lesssim M_\mathrm{Ch}=1.457\,(2Y_e)^2\ M_\odot\,,
\label{eq:chmass}
\end{equation}
where $Y_e = n_e/n_b$ is the number of electrons (number density
$n_e$) per baryon (number density $n_b = n_p + n_n$). Since $Y_e$
drops from an initial value around 0.46 for iron-group matter to
less than 0.3 after trapping, the homologous core shrinks to 
roughly 0.5\,$M_\odot$ \citep{Janka2012b}. 

\runinhead{(3) Core Bounce and Shock Formation}
Within milliseconds after trapping, corresponding to the free-fall 
time,
\begin{equation}
t_\mathrm{ff}\sim \frac{1}{\sqrt{G\rho}}\sim 
\frac{0.004}{\sqrt{\rho_{12}}}\ \mathrm{s} 
\label{eq:freefalltime}
\end{equation}
($G$ being the gravitational constant and $\rho_{12}$ the 
density in 10$^{12}$\,g\,cm$^{-3}$), the center reaches nuclear
matter density, where the heavy nuclei dissolve in a 
phase transition to a uniform nuclear medium. A sharp rise
of the incompressibility due to repulsive contributions to
the nuclear force between the nucleons provides resistance 
against further compression, and the collapse of the homologous 
inner core comes to an abrupt halt. As it bounces
back\index{core bounce}, sound 
waves steepen into a shock front\index{shock front}
at the boundary to the supersonically infalling outer layers
(Fig.~\ref{fig:janka-nuphases}, middle left panel). The bounce
shock\index{shock!bounce shock} 
begins to travel outwards against the ongoing collapse of
the overlying iron-core material.

\runinhead{(4) Shock Propagation and $\nu_e$ Burst at Shock Breakout}
Electron neutrinos are produced in huge numbers by electron 
captures on free protons behind the outward moving shock front.
However, they stay trapped in the dense postshock matter until
the shock reaches sufficiently low densities for the $\nu_e$
to diffuse faster than the shock propagates. At this moment,
the so-called shock breakout\index{shock!shock breakout} into 
neutrino-transparent layers, a luminous flash of $\nu_e$
--the breakout burst\index{breakout burst}-- is 
emitted (Fig.~\ref{fig:janka-nuphases}, middle right panel;
Sect.~\ref{sec:janka-nueburst}).

Shortly after shock breakout, the dramatic loss of $\nu_e$ 
leads to a considerable drop of the electron-lepton number 
in the shock-heated matter. This allows for the appearance of
large concentrations of positrons. Because of that and the 
compressional heating of the settling proto-neutron 
star\index{proto-neutron star}, which begins to assemble 
around the center, pair-production processes (mainly 
$e^-e^+$ pair annihilation and nucleon bremsstrahlung;
Table~\ref{tab:janka-nuprocesses} and 
Fig.~\ref{fig:janka-feynman}) become efficient and start
to create heavy-lepton neutrinos and antineutrinos. With 
positrons and neutrons becoming more and more abundant,
$e^+$ captures on neutrons also accomplish the emission of 
$\bar\nu_e$.

\runinhead{(5) Shock Stagnation and Revival by Neutrino Heating}
The shock front\index{shock front} is a sharp flow discontinuity
(whose narrow width is determined by the small, microphysical
viscosity of the stellar plasma), in which the kinetic energy
of the supersonically infalling preshock matter is dissipated
into thermal energy, leading to an abrupt deceleration and 
compression of the flow and a corresponding increase of the
density, temperature, pressure, and entropy behind the shock. 
Because of the temperature increase, heavy nuclei in the
preshock medium are disintegrated essentially completely to
free nucleons when the matter passes the shock. This consumes
appreciable amounts of energy, roughly 8.8\,MeV per nucleon or
$1.7\times 10^{51}$\,erg per 0.1\,$M_\odot$. This energy drain
and the additional energy losses by the $\nu_e$-burst reduce
the postshock pressure and weaken the expansion of the bounce
shock. It finally stagnates at a radius of typically less than
150\,km and an enclosed mass of around 1\,$M_\odot$, which is
still well inside the collapsing stellar iron core. The 
prompt bounce-shock mechanism therefore fails to initiate the
explosion of the dying star as supernova.

%----------------------------------------------------------------
\begin{figure}[t!]
\sidecaption[t]
\includegraphics[scale=0.41]{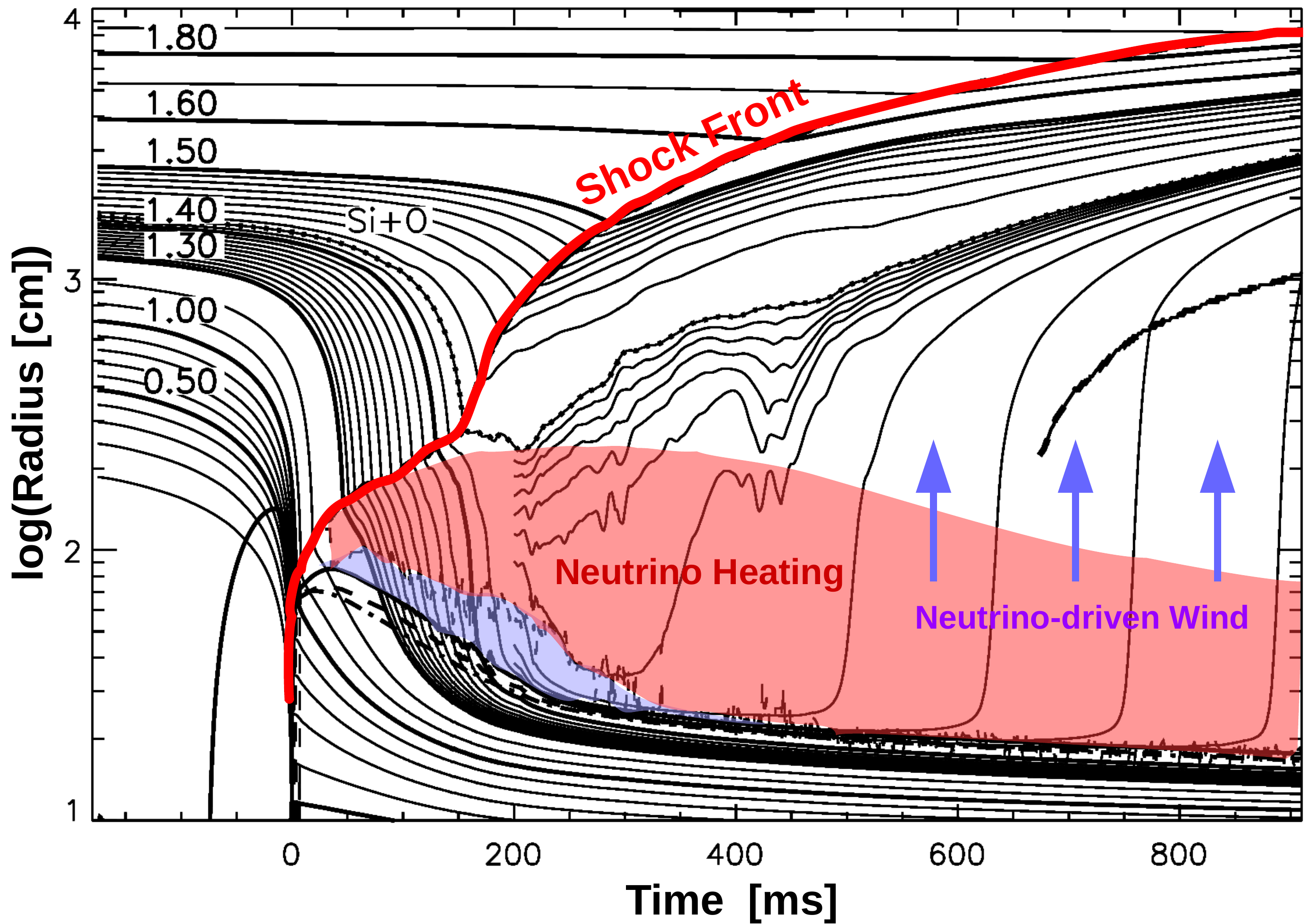}
\caption{Time evolution of the neutrino-driven explosion of a
15\,$M_\odot$ star as obtained in a multi-dimensional
hydrodynamic simulation, visualized by a mass-shell plot. The
horizontal axis shows time in milliseconds and the vertical
axis the radial distance (in cm) on a logarithmic scale. The black,
solid lines starting at the left edge of the plot belong to the
radii that enclose selected values of baryonic mass, in some cases
indicated by labels (in units of solar masses) next to the lines.
The line with overlaid crosses marks the boundary between the
silicon layer and the silicon-enriched oxygen layer of the
progenitor star. Retreating lines signal the collapse
of stellar shells and outgoing lines the expansion of matter
expelled in the beginning supernova explosion. The thick red line
marks the supernova shock front, which is formed at the moment
of core bounce (here chosen to define time $t = 0$).
The neutron star assembles from the
mass shells settling in the lower part of the image at $t > 0$.
The thick, black solid, dashed and dash-dotted lines that first
expand and then contract with the neutron star represent
the radial locations of the average neutrinospheres of $\nu_e$,
$\bar\nu_e$, and heavy-lepton neutrinos, respectively,
close to the surface
of the nascent neutron star. The light blue and red areas
denote the regions of neutrino cooling and neutrino heating,
respectively, outside of the neutrinospheres, which are
separated by the ``gain radius''\index{gain radius} (thin, 
dashed black line). The neutrino-driven
wind\index{neutrino-driven wind} (indicated
by blue arrows) is visible by mass shells that start their
outward expansion just above the neutron star surface. The
thick-thin dashed line beginning at about 700\,ms is the
wind-termination
shock\index{neutrino-driven wind!termination shock}
that is formed when the fast wind
collides with the slower preceding ejecta.
\citep[Figure adapted from][]{Pruet2005}
}
\label{fig:janka-s15shells}
\end{figure}
%----------------------------------------------------------------

The most likely mechanism to revive the stalled shock front
and to initiate its expansion against the ram pressure of the
collapsing surrounding stellar core matter, is energy transfer
by the intense neutrino flux radiated from the nascent
neutron star. The most important reactions for depositing 
fresh energy behind the shock are $\nu_e$ and $\bar\nu_e$
captures on free nucleons:
\begin{eqnarray}
\nu_e + n &\longrightarrow& p + e^- \,,  \label{eq:nueabs} \\
\bar\nu_e + p &\longrightarrow& n + e^+ \,.  \label{eq:barnueabs} 
\end{eqnarray}
Current numerical simulations, recently also performed in all
three spatial dimensions, demonstrate the viability of this
neutrino-heating mechanism\index{neutrino-heating mechanism} 
in principle 
(Fig.~\ref{fig:janka-s15shells}) so that this mechanism
appears as the most promising explanation of the far majority
of ``normal'' core-collapse supernovae. For stars more massive
than $\sim$10\,$M_\odot$, non-radial hydrodynamic instabilities
(like convective overturn\index{convective overturn}
and the standing accretion-shock
instability (SASI)\index{standing accretion-shock instability 
(SASI)}) provide
crucial support for the onset of the explosion, and also for stars
near the lower mass end of supernova progenitors 
($\sim$9--10\,$M_\odot$), non-spherical
flows play an important role for determining the energy and 
asymmetries of the explosion (see the indication of non-radial
mass motions in the left bottom panel of 
Fig.~\ref{fig:janka-nuphases}). Despite the promising results
of current models, many questions remain to be settled, and 
an ultimate confirmation of the neutrino-driven
mechanism\index{neutrino-driven mechanism} will 
require observational evidence. A high-statistics measurement 
of the neutrino signal from a future galactic supernova could
be a milestone in this respect.

Before the supernova shock front re-accelerates outward and the 
supernova blast is launched, stellar matter collapsing through
the stagnant shock feeds a massive accretion 
flow\index{proto-neutron star!accretion} onto the 
nascent neutron star (typically several 0.1\,$M_\odot$\,s$^{-1}$).
The hot accretion mantle\index{proto-neutron star!accretion mantle}
around the high-density, lower-entropy 
core of the neutron star radiates high fluxes mainly of $\nu_e$ and
$\bar\nu_e$, which carry away the 
gravitational binding energy\index{gravitational binding energy}
that is released in the gravitational collapse. This accretion
luminosity adds to the core luminosity of all species of 
neutrinos and antineutrinos ($\nu_e$, $\bar\nu_e$, and $\nu_x$)
that diffuse out from the deeper layers
(Fig.~\ref{fig:janka-nuphases}, bottom left panel).

\runinhead{(6) Proto-neutron Star Cooling and Neutrino-driven Wind}
Accretion does not subside immediately after the explosion sets
in. There can be an extended phase of continued mass accretion
by the nascent neutron star that proceeds simultaneously to the 
outward acceleration of mass behind the outgoing shock. Eventually,
however, after hundreds of milliseconds up to maybe a second,
depending on the progenitor star and the speed of shock expansion,
accretion ends and the proto-neutron
star\index{proto-neutron star} enters its 
Kelvin-Helmholtz cooling
phase\index{Kelvin-Helmholtz cooling}, in which it loses its remaining
gravitational binding energy\index{gravitational binding energy}
by the emission of neutrinos and 
antineutrinos of all flavors on the time scale of neutrino 
diffusion. Based on a simple diffusion model for a homogeneous
sphere, \citet{Burrows1984,Burrows1990} derived order-of-magnitude
estimates for the deleptonization and energy-loss time scales:
\begin{eqnarray}
t_\mathrm{L} &\sim& \frac{3 R_\mathrm{ns}^2}{\pi^2c\lambda_0}\,
\frac{\mathrm{d}Y_\mathrm{L}}{\mathrm{d}Y_{\nu_e}} \sim 3\ \mathrm{s}
\,,
\label{eq:tdelep} \\
t_\mathrm{E} &\sim& \frac{3 R_\mathrm{ns}^2}{\pi^2c\lambda_0}\,
\frac{E_\mathrm{th}^0}{2 E_\nu^0} \sim 10\ \mathrm{s} \,,
\label{eq:terg} 
\end{eqnarray}
where $R_\mathrm{ns}\approx 10$\,km is the neutron-star radius,
\begin{equation}
\lambda_0 = \frac{1}{n_b\langle\sigma_\nu\rangle}
\sim 10\,\,\mathrm{cm}\,\,
\left(\frac{E}{100\,\mathrm{MeV}}\right)^{\! -2}
\left(\frac{M_\mathrm{ns}}{1.5\,M_\odot}\right)^{\! -1}
\left(\frac{R_\mathrm{ns}}{10\,\mathrm{km}}\right)^{\! 3} 
\label{eq:lambda0}
\end{equation}
the initial average mean free path of the neutrinos,
$E_\mathrm{th}^0$ and $E_\nu^0$ the initial total baryon and 
neutrino thermal energies, respectively, and the 
ratio of these thermal energies as well as
$\mathrm{d}Y_\mathrm{L}/\mathrm{d}Y_{\nu_e}$
describe the ability of the neutron star to
replenish the loss of lepton number and energy due to the radiated
neutrinos from the available reservoirs of these quantities.
While the temperatures in the interior of the newly formed
neutron star can reach up to more than 50\,MeV and the thermal
energies of neutrinos can be 100\,MeV and higher, these 
high-energy neutrinos are absorbed, re-emitted and 
downscattered billions of times before they escape from
the neutrinospheric region with final mean energies of
10--20\,MeV over much of the Kelvin-Helmholtz
phase\index{Kelvin-Helmholtz cooling}.

While the proto-neutron star deleptonizes and cools by neutrino
losses, the energetic neutrinos radiated from the neutrinosphere
continue to deposit energy in the overlying, cooler layers,
mainly by the reactions of Eqs.~(\ref{eq:nueabs}) and 
(\ref{eq:barnueabs}). This leads to a persistent, dilute
outflow of mass (with initial mass-loss rates of typically
several $10^{-2}$\,$M_\odot$\,s$^{-1}$) from the surface
of the nascent neutron star. This so-called neutrino-driven
wind\index{neutrino-driven wind} (Fig.~\ref{fig:janka-nuphases}, 
bottom right panel, and Fig.~\ref{fig:janka-s15shells}) is 
discussed as potential site for the formation of
trans-iron elements. The mass-loss rate,
entropy, and expansion velocity of this wind are sensitive
functions of the neutron-star radius and mass and of the
luminosities and spectral hardness of the emitted neutrinos 
\citep{Qian1996,Otsuki2000,Thompson2001,Arcones2007}.
Even more important is the fact that the 
neutron-to-proton ratio of the expelled matter is determined
by the luminosity and spectral differences of $\nu_e$ and 
$\bar\nu_e$, which leads to an interesting sensitivity of
the nucleosynthetic potential of this environment to the
nuclear physics of the neutron star medium and to non-standard
neutrino physics like flavor oscillations or the speculative
existence of sterile neutrinos.

%----------------------------------------------------------------------
\begin{figure}[t]
\sidecaption[t]
\includegraphics[scale=.42]{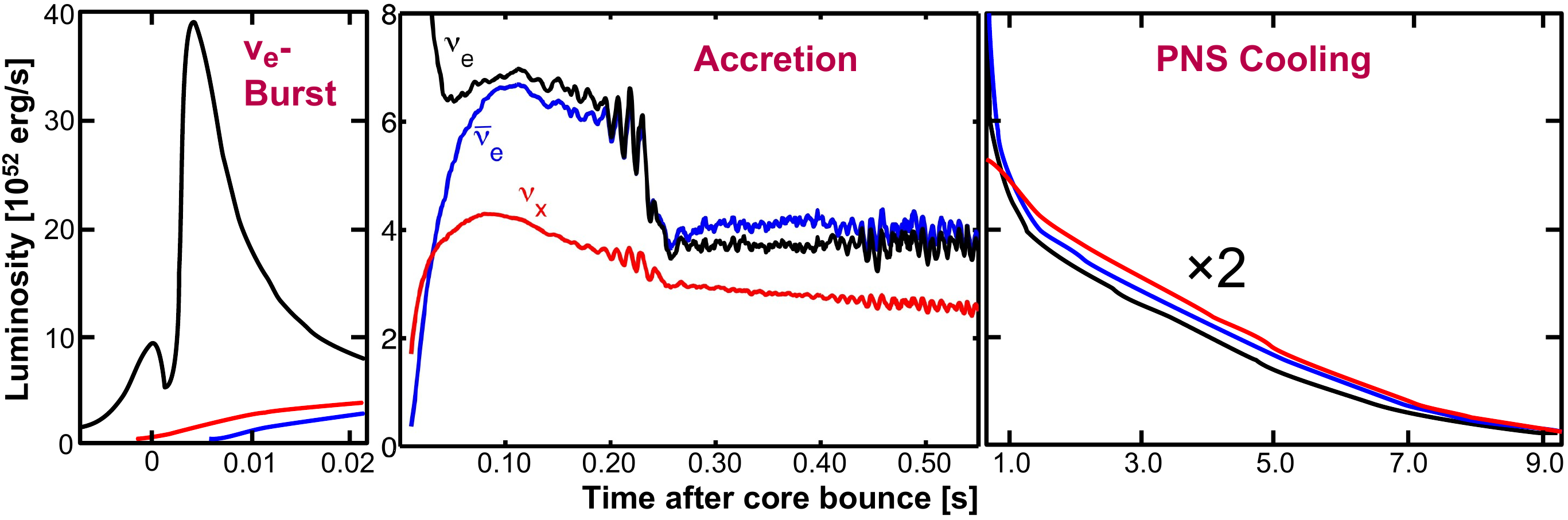}
\caption{Neutrino luminosities ($\nu_e$: black; $\bar\nu_e$: blue;
$\nu_x$ as one species of $\nu_\mu$, $\bar\nu_\mu$, $\nu_\tau$,
$\bar\nu_\tau$: red) during the main neutrino-emission
phases. The {\em left panel} shows the prompt burst of electron
neutrinos associated with the moment of shock breakout into the
neutrino-transparent outer core layers only milliseconds after
bounce ($t = 0$). The {\em middle panel} corresponds to the post-bounce
accretion phase before shock revival as computed in a three-dimensional
simulation \citep[see][]{Tamborra2014}. The quasi-periodic luminosity
variations are a consequence of modulations of the mass-accretion rate
by the neutron star caused by violent non-radial motions due to
hydrodynamic instabilities (in particular due to the standing 
accretion-shock instability 
or SASI)\index{standing accretion-shock instability (SASI)} in the
postshock layer. The {\em right panel} displays the decay of the
neutrino luminosities over several seconds in the neutrino-cooling
phase of the newly formed neutron star (the plotted values are
scaled up by a factor of 2)
}
\label{fig:janka-nulums}
\end{figure}
%-------------------------------------------------------------------------

%---------------------------------------------------------------------------------------
\begin{figure}[t]
\sidecaption[t]
\includegraphics[scale=.39]{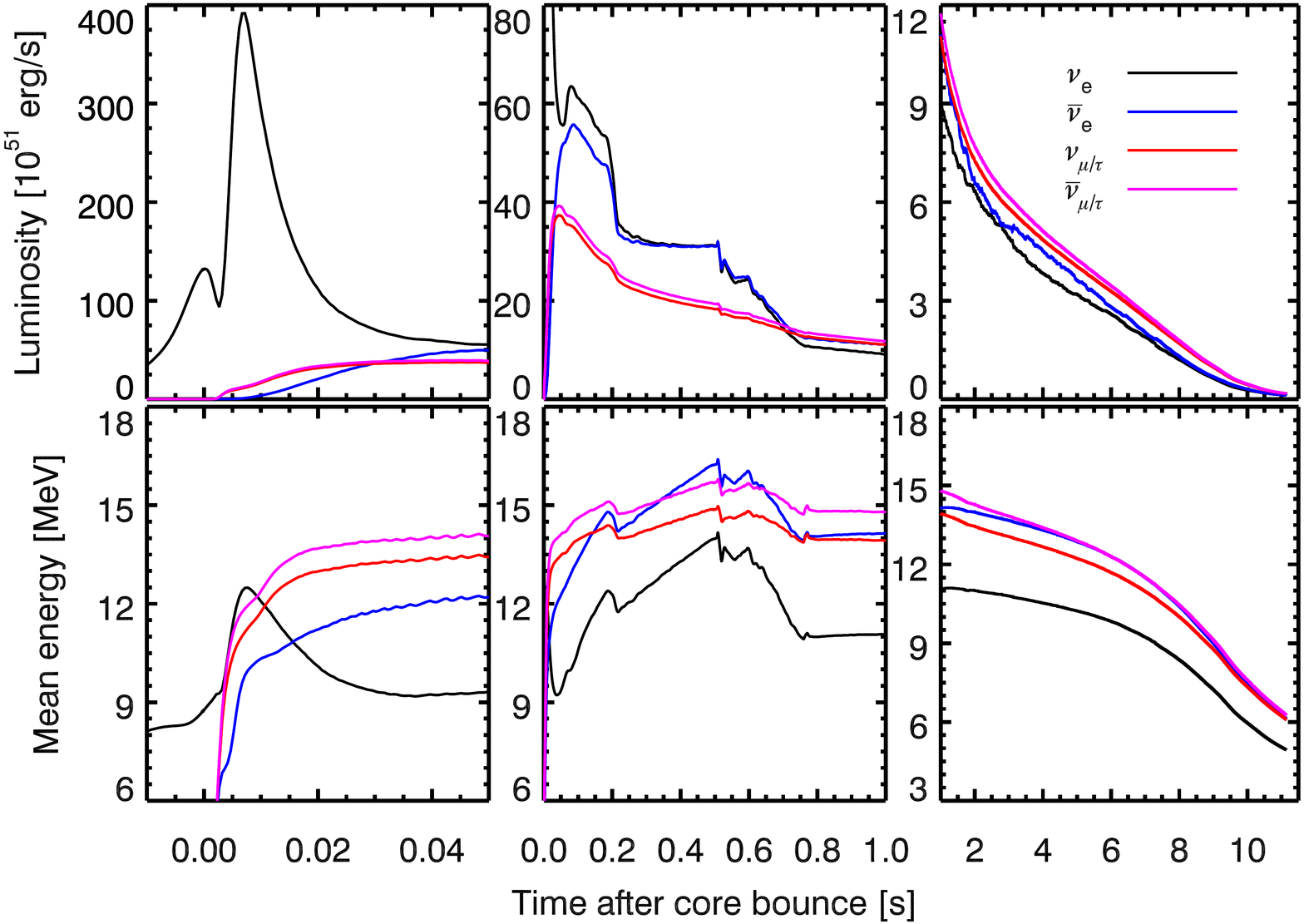}
\caption{
Neutrino signal computed for the supernova explosion of a star of
27\,$M_\odot$, which gives birth to a neutron star with 1.6\,$M_\odot$.
The {\em left panels} correspond to the shock-breakout phase, the {\em middle panels}
to the post-bounce accretion phase including the transition to the proto-neutron
star cooling phase, which is given in the {\em right panels}.
The {\em upper panels} display the neutrino luminosities
($\nu_e$ black; $\bar\nu_e$: blue; one species of $\nu_{\mu,\tau}$: red;
one species of $\bar\nu_{\mu,\tau}$: magenta), 
and the {\em lower panels} panels display the
mean energies of the radiated neutrinos. In contrast to Fig.~\ref{fig:janka-nulums},
the differences of heavy-lepton neutrinos and antineutrinos associated with
weak-magnetism corrections of neutrino-nucleon scattering are shown. The
slightly lower scattering opacity of $\bar\nu_{\mu,\tau}$ leads to slightly
higher luminosities and higher mean energies (by $\sim$1\,MeV) compared to those
of $\nu_{\mu,\tau}$.
The explosion sets in at 0.5\,s after core bounce, but accretion onto the
proto-neutron star ends only at about 0.75\,s, which marks the onset of the
cooling phase. (Figure courtesy of Robert Bollig)
}
\label{fig:janka-lums27}
\end{figure}
%----------------------------------------------------------------------------------------

\section{Neutrino Emission Properties}
\label{sec:janka-properties}

Three main phases of neutrino emission can be discriminated that 
correspond to the dynamical evolution stages described in the previous
section \citep{Janka1993}. They are
displayed in Fig.~\ref{fig:janka-nulums} and described in the 
following three subsections.

\subsection {Shock-breakout Burst of Electron Neutrinos}
\label{sec:janka-nueburst}

A luminous
flash of neutronization neutrinos is radiated when the shock 
transitions from the opaque to the neutrino-transparent,
low-density ($\rho\lesssim 10^{11}$\,g\,cm$^{-3}$)
outer layers of the iron core. At this moment, typically setting in 
$\sim$2\,ms after core bounce, the large number of $\nu_e$ created 
by electron captures on free protons in the shock-heated matter
can ultimately escape. During the
preceding collapse prior to core bounce, the $\nu_e$ emission rises
continuously because an increasingly bigger fraction of the stellar
core is compressed to densities where efficient electron captures
become possible. Only within a brief period ($\pm$1\,ms) around core
bounce, the strong compression and Doppler redshifting of the main
region of $\nu_e$ generation lead to a transient dip in the $\nu_e$
luminosity. At shock breakout, also the luminosities of heavy-lepton 
neutrinos and shortly afterwards those of $\bar\nu_e$ begin to rise,
because their production by pair processes becomes possible in the
shock-heated matter (see Sect.~\ref{sec:janka-phases};
Fig.~\ref{fig:janka-nulums}, left panel). The $\nu_e$
luminosity burst and the rise phase of the $\bar\nu_e$ and $\nu_x$
luminosities show a generic behavior with little dependence on the
progenitor star \citep{Kachelriess2005}. The burst reaches
a peak luminosity near $4\times 10^{53}$\,erg\,s$^{-1}$,
has a half-width of less than 10\,ms and 
releases about $2\times 10^{51}$\,erg of energy
within only 20\,ms. The mean energy of the radiated $\nu_e$ 
also peaks at the time of maximum luminosity and reaches
12--13\,MeV (Figure~\ref{fig:janka-lums27}, lower left panel).

\subsection{Post-bounce Accretion} 

This phase follows when the 
$\nu_e$ luminosity declines from the maximum and levels off into
a plateau. Both $\nu_e$ and $\bar\nu_e$ are produced in large 
numbers by charged-current processes in the hot mantle of the 
proto-neutron star. The mass of this mantle grows continuously,
because it is fed by the accretion flow of the collapsing stellar
matter that falls through the stagnant shock and is heated by
compression. The luminosities of $\nu_e$ and $\bar\nu_e$ are
very similar during this phase with a slight number excess of
$\nu_e$ because of ongoing deleptonization. In contrast, the
individual luminosities of $\nu_x$ are considerably lower. These
neutrinos originate mostly from the denser core region, where
the high densities and temperatures allow nucleon bremsstrahlung
to generate $\nu_x\bar\nu_x$ pairs.

The neutrino emission (luminosities and mean energies) during 
the accretion phase show large variations between different 
progenitor stars, because the neutrino quantities 
scale with the mass accretion 
rate, $\dot M(t)$, and the growing proto-neutron star mass, 
$M_\mathrm{ns}(t)$. Both $\dot M$ and $M_\mathrm{ns}$ are
higher for progenitor stars that possess
more compact cores\index{core compactness}, i.e., where a 
certain mass is condensed into a smaller volume prior to collapse.
Progenitors with higher core compactness (which tend to be
more massive, too, but with considerable non-monotonic variations)
therefore radiate higher luminosities and harder neutrino spectra
\citep{Janka2012b,OConnor2013}.

Moreover, non-radial flows in the supernova core, which are
a consequence of hydrodynamic
instabilities\index{hydrodynamic instability} in the proto-neutron
star\index{proto-neutron star}
and in the region behind the stalled shock front
(like convective overturn\index{convective overturn}
and the standing accretion-shock
instability (SASI)\index{standing accretion-shock instability
(SASI)}), can cause
large-scale temporal modulations of the accretion flow onto the 
neutron star. This can lead to 
time- and direction dependent, large-amplitude, quasi-periodic 
fluctuations of the luminosities and mean energies of the radiated 
neutrinos during the accretion phase 
\citep[Fig.~\ref{fig:janka-nulums}, 
middle panel][]{Lund2012,Tamborra2013,Tamborra2014}.

The instantaneous spectra of the radiated muon and tau neutrinos
are reasonably well described by Fermi-Dirac
functions\index{Fermi-Dirac distribution} with zero 
degeneracy, and their luminosities can be expressed by a 
Stefan-Boltzmann like formula as
\begin{equation}
L_{\nu_x} = 4\pi \phi s_\nu R_\mathrm{ns}^2 T_\nu^4 \,,
\label{eq:lnux}
\end{equation}
where the average energy and the 
effective spectral temperature $T_\nu$ (measured in MeV)
are linked by $\langle E\rangle = 3.15\,T_\nu$. 
$R_\mathrm{ns}$ is the radius of the proto-neutron star and
$s_{\nu}=4.50\times 10^{35}$\,erg\,MeV$^{-4}$cm$^{-2}$s$^{-1}$
for a single species of $\nu_x$.
The ``greyness factor'' $\phi$ is determined by numerical
simulations to range between $\sim$0.4 and $\sim$0.85
\citep{Mueller2014}.

Since the emission of $\nu_e$ and $\bar\nu_e$ is
enhanced by the accretion component, the sum of their 
luminosities can be written as
\begin{equation}
L_{\nu_e} + L_{\bar\nu_e} = 2\beta_1 L_{\nu_x} +
\beta_2\,\frac{G M_\mathrm{ns}\dot M}{R_\mathrm{ns}} \,.
\label{eq:acclum}
\end{equation}
The first term on the r.h.s.\ represents the ``core component''
of the luminosity carried by neutrinos diffusing out from the
high-density inner regions of the proto-neutron star. This
component can be assumed to be similar to the luminosity of 
$\nu_\mu$ plus $\bar\nu_\mu$,
because the core radiates all types of neutrinos in
roughly equal numbers from a reservoir in thermal equilibrium,
which is confirmed by the close similarity of the luminosities
of all neutrino species after the end of accretion.
The second term on the r.h.s.\ stands for the accretion
component expressed by the product of mass accretion
rate, $\dot M$, and Newtonian surface gravitational potential
of the neutron star, $\Phi = GM_\mathrm{ns}/R_\mathrm{ns}$.
By a least-squares fit to a large set of 1D results for the
post-bounce accretion phase of different progenitor stars, 
values between $\beta_1\approx 1.25$ and $\beta_2\approx 0.5$
can be deduced (L.~H\"udepohl, 2014, private communication), 
which depend only weakly on the nuclear EoS.
The values apply later than about 150\,ms after bounce, when the
postshock accretion layer has settled into a quasi-steady
state. \citet{Mueller2014} used a form slightly
different from Eq.~(\ref{eq:acclum}) with $\beta_1 = 1$;
they found $\beta_2\approx 0.5$--1 prior to explosion.

During the accretion phase the mean energies of all neutrino
species show an overall trend of increase, which is
typically steeper for $\nu_e$ and $\bar\nu_e$ than for $\nu_x$.
The secular rise of the mean energies of the radiated
$\nu_e$ and $\bar\nu_e$ is fairly
well captured by the proportionality $\langle E_\nu\rangle
\propto M_\mathrm{ns}(t)$. The proportionality constant
depends on the neutrino type but
is only slightly progenitor-dependent \citep{Mueller2014}.
This secular rise of $\langle E_{\nu_e}\rangle$ and
$\langle E_{\bar\nu_e}\rangle$ is supported by a 
local temperature maximum somewhat
inside of the neutrinospheres of these neutrinos, which
forms because of compressional heating of the growing accretion
layer in progenitors with sufficiently high accretion rates
(typically more massive than about 10\,$M_\odot$).
Because of the continuous growth of the mean energies with
$M_\mathrm{ns}(t)$, the canonical order of the average energies,
$\left<E_{\nu_e}\right> < \left<E_{\bar\nu_e}\right>
< \left<E_{\nu_x}\right>$ changes (transiently) to
$\left<E_{\nu_e}\right> < \left<E_{\nu_x}\right>
< \left<E_{\bar\nu_e}\right>$ (Fig.~\ref{fig:janka-lums27},
lower middle panel). 
This hierarchy inversion is enhanced and shifted to
earlier times when energy transfer in neutrino-nucleon scattering
is taken into account. Non-isoenergetic neutrino-nucleon scattering
reduces the mean energies of $\nu_x$ in the ``high-energy filter''
layer between the $\nu_x$ energy sphere and the $\nu_x$ transport
sphere \citep[see Sect.~\ref{sec:janka-decoupling} and
Fig.~\ref{fig:janka-nutrans}][]{Raffelt2001,Keil2003}.
The corresponding
energy transfer to the stellar medium also raises the luminosities
and mean energies of $\nu_e$ and $\bar\nu_e$.
Different from the mean energies, the mean squared energies,
$\left<E_\nu^2\right>$, and rms energies always
obey the canonical hierarchy.

\subsection{Kelvin-Helmholtz Cooling and Deleptonization of the
Proto-neutron Star}  

After the explosion has set in, the proto-neutron
star continues to radiate lepton number and energy by high 
neutrino fluxes for many seconds (Sect.~\ref{sec:janka-phases}).
The luminosities of all kinds of neutrinos and antineutrinos become
similar (within $\sim$10\%) during this phase
and decline with time in parallel
(Fig.~\ref{fig:janka-nulums}, right panel). The typical average
luminosities during Kelvin-Helmholtz
cooling\index{Kelvin-Helmholtz cooling} are of the order of
\begin{equation}
L_\nu^\mathrm{tot} \equiv \sum_{i = e,\mu,\tau} L_{\nu_i}+L_{\bar\nu_i}
\sim \frac{E_\mathrm{b}}{t_\mathrm{E}} \sim 
\mathrm{several}\ 10^{52}\ \mathrm{erg\,s}^{-1} \,.
\label{eq:ldiff}
\end{equation}
Rough estimates of $E_\mathrm{b}$ and $t_\mathrm{E}$ 
were provided by Eq.~(\ref{eq:ebind}) and
Eq.~(\ref{eq:terg}), respectively. Around
about 1\,second, the mean energies of the radiated neutrinos show
a turnover and begin to decrease, reflecting the gradual cooling 
of the outer layers of the proto-neutron star
(Fig.~\ref{fig:janka-lums27}). A thick convective shell
inside the star grows in mass while its inner boundary progresses 
towards the center \citep[see][]{Mirizzi2015}. 
Convective energy transport in the high-density core of the 
neutron star is faster than diffusive transport and considerably
accelerates the lepton number and energy loss through neutrinos.
Because of the ``high-energy filter'' effect of the extended 
scattering atmosphere between the energy and transport spheres of
heavy-lepton neutrinos, $\nu_{\mu,\tau}$ with their 
higher scattering opacity are radiated with slightly softer
spectra than $\bar\nu_e$ and $\bar\nu_{\mu,\tau}$ during all
of the Kelvin-Helmholtz cooling
evolution\index{Kelvin-Helmholtz cooling}
(Fig.~\ref{fig:janka-lums27}, lower right panel).

%----------------------------------------------------------------------------------------
\begin{figure}[t]
\sidecaption[t]
\includegraphics[scale=.50]{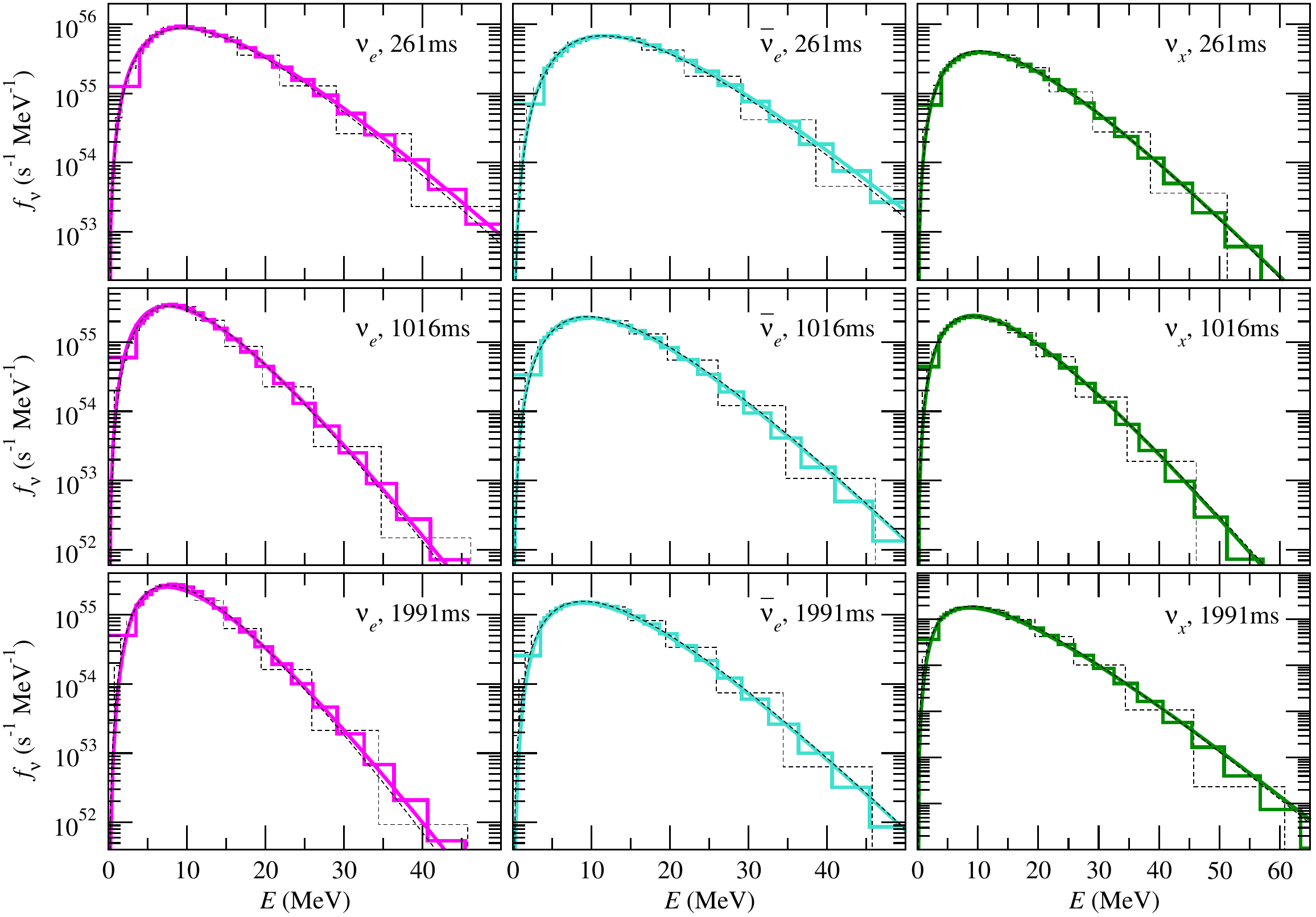}
\caption{
Spectra for electron neutrinos ($\nu_e$; {\em left column}), electron antineutrinos
($\bar{\nu}_e$; {\em middle column}), and heavy-lepton neutrinos ($\nu_x$, {\em right
column}) during the accretion phase (261\,ms after core bounce, {\em top row}) and
for two times during the proto-neutron star cooling phase (1016\,ms, {\em middle
row}; 1991\,ms, {\em bottom row}). The step functions are results of numerical
simulations with lower (thin dashed) and higher (thick, colored) resolution.
The continuous curves are quasi-thermal fits according to
Eq.~(\ref{eq:fitspectrum}) for the lower resolution (thin dashed lines) and
higher resolution (thick solid lines) cases. All $\alpha$ values for the fit
functions are in the interval $2.3\le \alpha\le 3.3$.
\citep[Figure taken from][]{Tamborra2012}
}
\label{fig:janka-spectra}
\end{figure}
%----------------------------------------------------------------------------------------

\subsection{Spectral Shape}

The spectra of radiated neutrinos are usually somewhat different
from thermal spectra. Since neutrino-matter interactions are
strongly energy dependent, neutrinos of different
energies decouple from the background medium at different radii
with different temperatures of the stellar plasma. Nevertheless,
the emitted neutrino spectrum\index{neutrino spectrum} 
can still be fitted by a Fermi-Dirac
distribution\index{Fermi-Dirac distribution},
$f(E) \propto E^2[1+\exp{(E/T-\eta)}]^{-1}$,
with fit temperature $T$ (in energy units) and effective 
degeneracy parameter $\eta$ \citep{Janka1989}.
A mathematically more convenient representation was introduced
by \citet{Keil2003}, who proposed the following
dimensionless form for the energy distribution at a large
distance from the radiating source:
\begin{equation}
f_\alpha(E)\propto \left(\frac{E}{\langle E\rangle}\right)^{\! \alpha}
e^{-(\alpha+1)E/\langle E\rangle} \,,
\label{eq:fitspectrum}
\end{equation}
where
\begin{equation}
\langle E\rangle = 
\frac{\int_0^\infty\mathrm{d}E\,E 
f_\alpha(E)}{\int_0^\infty\mathrm{d}E\,f_\alpha(E)}
\label{eq:meane}
\end{equation}
is the average energy.
The parameter $\alpha$ represents the amount of spectral
``pinching''\index{neutrino spectrum!pinching}
and can be computed from the two lowest energy
moments of the spectrum, $\langle E\rangle$ and
$\langle E^2\rangle$, by
\begin{equation}
\frac{\langle E^2\rangle}{\langle E\rangle^2}
= \frac{2+\alpha}{1+\alpha} \,.
\label{eq:alpha}
\end{equation}
Higher energy moments $\langle E^\ell\rangle$ for $\ell > 1$
are defined analogue to Eq.~(\ref{eq:meane}) with $E^\ell$
under the integral in the numerator instead of $E$.
Besides its analytic simplicity, this functional form has the
advantage to also allow for the representation of a wider range
of values for the spectral (anti-)pinching than a Fermi-Dirac fit.
A Fermi-Dirac spectrum with vanishing degeneracy parameter
($\eta=0$) corresponds to $\alpha\approx 2.3$,
a Maxwell-Boltzmann spectrum to $\alpha=2$, and
$\alpha \gtrsim 2.3$ yields ``pinched'' 
spectra\index{neutrino spectrum!pinching} (i.e.,
narrower than a thermal Fermi-Dirac spectrum), whereas
$\alpha \lesssim 2.3$ gives anti-pinched ones.
Comparing to high-resolution transport results,
\citet{Tamborra2012} showed that these 
``$\alpha$-fits''\index{neutrino spectrum!alpha-fit}
also reproduce the high-energy tails of the radiated
neutrino spectra very well (Fig.~\ref{fig:janka-spectra}).

The shape parameter $\alpha$ is up to 6--7 for $\nu_e$ around
the $\nu_e$-burst, and in the range of 2--3 for all neutrino
species at times later than $\sim$200\,ms after bounce
\citep{Mirizzi2015}.
In particular $\nu_e$ and $\bar\nu_e$ exhibit a tendency of
pinched spectra. This spectral pinching can be understood 
as a consequence of the energy dependence of the neutrino 
interactions and can be exemplified by considering the radiated 
luminosity spectrum as a combination of thermal contributions
from different, energy-dependent decoupling regions:
\begin{equation}
\frac{\mathrm{d}L_\nu(E)}{\mathrm{d}E} \approx
\frac{1}{4}c\,(4\pi R_\nu^2(E))\,B_\nu(E) \equiv
\pi\,c\,R_\nu^2(E)\,
\frac{4\pi}{(hc)^3}\,\frac{E^3}{1+\exp[(E-\mu_\nu)/T]}\,,
\label{eq:pinching}
\end{equation}
with $T = T(R_\nu(E))$ and $\mu_\nu = \mu_\nu(R_\nu(E))$ being
the gas temperature (in MeV) and neutrino equilibrium chemical 
potential at decoupling radius $R_\nu(E)$. The smaller
interaction cross section and opacity of low-energy neutrinos 
(Eq.~\ref{eq:sigma}) lead to their energetic decoupling at a
smaller radius (cf.\ Eq.~\ref{eq:esphere}), whereas 
high-energy neutrinos decouple at larger radii, where the 
stellar temperature is lower. These effects cause a 
reduction of the low-energy and high-energy wings on both
sides of the spectral peak compared to a thermal spectrum
with the temperature of the spectral maximum.

\section{Conclusions}
\label{sec:janka-conclusions}

Theoretical predictions of the neutrino emission from 
supernovae have become considerably more reliable and 
detailed since improved
transport treatments have become available in numerical
simulations after the change of the millennium. This Chapter
provides an overview of the foundations of the neutrino
physics in collapsing stars. Moreover, it presents
a summary of our current understanding of production and 
properties of the neutrino signal emitted during supernova
explosions and the birth of neutron stars.

The most advanced methods for describing neutrino transport 
in computational supernova models in spherical symmetry apply
solvers for the time-dependent Boltzmann transport
equation\index{Boltzmann transport equation}
\citep{Liebendoerfer2001,Liebendoerfer2004,Lentz2012}
or for the set of its first
two moment equations with a variable Eddington factor
closure derived from a simplified Boltzmann equation
\citep{Rampp2002,Mueller2010}. 
Both approaches take into account the velocity 
dependence of the neutrino transport, general relativistic 
effects, and the full phase-space dependence of the neutrino
interaction rates summarized in
Table~\ref{tab:janka-nuprocesses} and Fig.~\ref{fig:janka-feynman}.
These most advanced codes have been shown to yield results of
similar quality and overall consistency between each other
\citep{Liebendoerfer2005,Marek2006,Mueller2010}. 
These methods constitute the present
state-of-the-art for simulating neutrino transport in supernovae
and proto-neutron stars in spherical symmetry.

In three-dimensional supernova modeling similar 
sophistication is not yet feasible.
The current forefront here is defined by ray-by-ray
implementations of the two-moment method with Boltzmann closure 
\citep{Buras2006,Melson2015a,Melson2015b} and of flux-limited 
diffusion \citep{Lentz2015}, and the application of
two-moment schemes with algebraic closure relations is in sight.
Solving the time-dependent Boltzmann transport
problem in six-dimensional phase-space, however, is still on
the far horizon and remains a challenging task for future
supercomputing on the exascale level.

The same is true for a fully self-consistent inclusion of
the effects of neutrino flavor transformations. 
Matter-background induced oscillations according to the
Mikheyev-Smirnov-Wolfenstein (MSW) effect 
\citep{Wolfenstein1978,Mikheyev1985}
for the three active neutrino flavors occur 
at densities far below those of the supernova core (around
100\,g\,cm$^{-3}$ and at several 1000\,g\,cm$^{-3}$) and
must be taken into account when neutrinos propagate through
the dying star on their way to the terrestrial detector.
Since matter oscillations are suppressed in the dense interior
at conditions far away from the MSW resonances
\citep{Wolfenstein1979,Hannestad2000}, flavor
mixing inside of the neutrinosphere can be safely ignored,
unless fast pairwise neutrino conversions play a role 
\citep[][and references therein]{Izaguirreetal2017}.
Outside of the neutrinosphere, however, the neutrino densities
are so enormous that the large $\nu$-$\nu$ interaction 
potential can trigger self-induced flavor conversions. 
The possible consequences of this interesting effect have
so far been explored in post-processing studies using 
unoscillated neutrino data from numerical supernova 
simulations \citep[for a status report, see][]{Mirizzi2015}.
The highly complex and rich phenomenology of these
self-induced flavor changes, however, is not yet settled,
and therefore final conclusions on their possible effects 
for the supernova physics and for neutrino detection cannot
be drawn yet.

The detection of a high-statistics neutrino signal from a 
supernova in the Milky Way is a realistic possibility with
existing and upcoming experimental facilities. Such a
measurement will provide unprecedentedly detailed and direct
information of the physical conditions and of the dynamical
processes that facilitate and accompany the collapse and
explosion of a star and the formation of its compact 
remnant. A discovery of the diffuse supernova neutrino
background as integrated signal of all past stellar 
collapse events seems to be in reach \citep[for a review,
see][]{Mirizzi2015}.
It will put our fundamental understanding 
of the neutrino emission from the whole variety of stellar
death events to the test and may offer the potential to
set constraints on neutron star versus black hole formation
rates.

\runinhead{R{\'e}sum{\'e}}
Neutrinos are crucial agents during all stages of 
stellar collapse and explosion. Besides gravitational
waves\index{gravitational waves} 
they are the only means to obtain direct
information from the very heart of dying stars.
Therefore they are a unique probe of the physics that 
plays a role at extreme conditions that otherwise
are hardly accessible to laboratory experiments.
The total energy, luminosity evolution, spectral 
distribution, and the mix of different species, which
describe the radiated neutrino signal, carry
imprints of the thermodynamic conditions, 
dynamical processes, and characteristic properties
of the progenitor star and of its compact remnant.
Numerical models are advanced to an increasingly
higher level of realism for better predictions of the
measurable neutrino features and their consequences.

\begin{acknowledgement}
The author is indebted to Georg Raffelt for valuable
discussions and thanks him and Robert Bollig for providing
graphics used in this article. Research by the author was
supported by the European Research Council through an
Advanced Grant (ERC-AdG No.\ 341157-COCO2CASA), by
the Deutsche Forschungsgemeinschaft through 
the Cluster of Excellence ``Universe'' (EXC-153), and
by supercomputing time from the European PRACE
Initiative and the Gauss Centre for Supercomputing.
\end{acknowledgement}
\section*{Cross-References}

{\em Neutrino-driven Explosions}\\
{\em Neutrinos and Their Impact on Core-Collapse Supernova Nucleosynthesis}\\
{\em Neutrino Conversion in Supernovae}\\
{\em Neutrino Signatures from Young Neutron Stars}\\
{\em Explosion Physics of Core-Collapse Supernovae}\\
{\em Neutrinos from Core-Collapse Supernovae and Their Detection}\\
{\em The Diffuse Neutrino Flux from Supernovae}\\
{\em Gravitational Waves from Supernovae}

\bibliography{JankaReferences-file}

\end{document}